\documentclass[
aps,prl,reprint,amsmath,amssymb
]{revtex4-2}
\usepackage[T1]{fontenc}
\usepackage{lmodern,graphicx,bm,microtype}
\usepackage[colorlinks=true,allcolors=blue]{hyperref}
\newcommand{\dd}{\mathrm d}
\newcommand{\ee}{\mathrm e}
\newcommand{\tc}{\tau_{\mathrm C}}
\begin{document}
\title{Aggregating soft-photon recoil with a positive finite-step sampler}

\author{Nina Elkina$^1$ and Alexander Pukhov$^{1}$}
\affiliation{$^1$Institut f\"ur Theoretische Physik I, Heinrich-Heine-Universit\"at D\"usseldorf, 40225 D\"usseldorf, Germany}

\date{November 19, 2026}

\begin{abstract}
We model stochastic radiation reaction by aggregating soft photons into one
positive recoil, retaining the local no-emission probability and two loss
moments of the crossed-field QED kernel. Unlike Gaussian diffusion, the
sampler preserves one-sided energy loss. Analytical synchrotron moments
expose a finite-step cooling bias, which state feedback reduces. Laser
and ion-channel tests probe excessive-loss risk and directed radiation
spectra, providing a practical route to stochastic particle simulations
without resolving individual soft emissions.
\end{abstract}
\maketitle

\paragraph{Introduction.}
Radiation reaction can reshape electron motion and the radiation
produced in intense electromagnetic fields
\cite{Fedotov2023physrep,DiPiazza2012rmp,Cole2018prx,Poder2018prx,Wistisen2018}.
When most photons are soft, many emissions can occur before the
electron's state changes appreciably. This separation suggests a
coarse-grained description that retains recoil fluctuations without
resolving every emission individually. Accurate treatment of this
soft radiation is also relevant to background estimates for proposed
measurements of vacuum birefringence \cite{KingElkina2016pra}.

\begin{figure*}[t]
\centering\includegraphics[width=\textwidth]{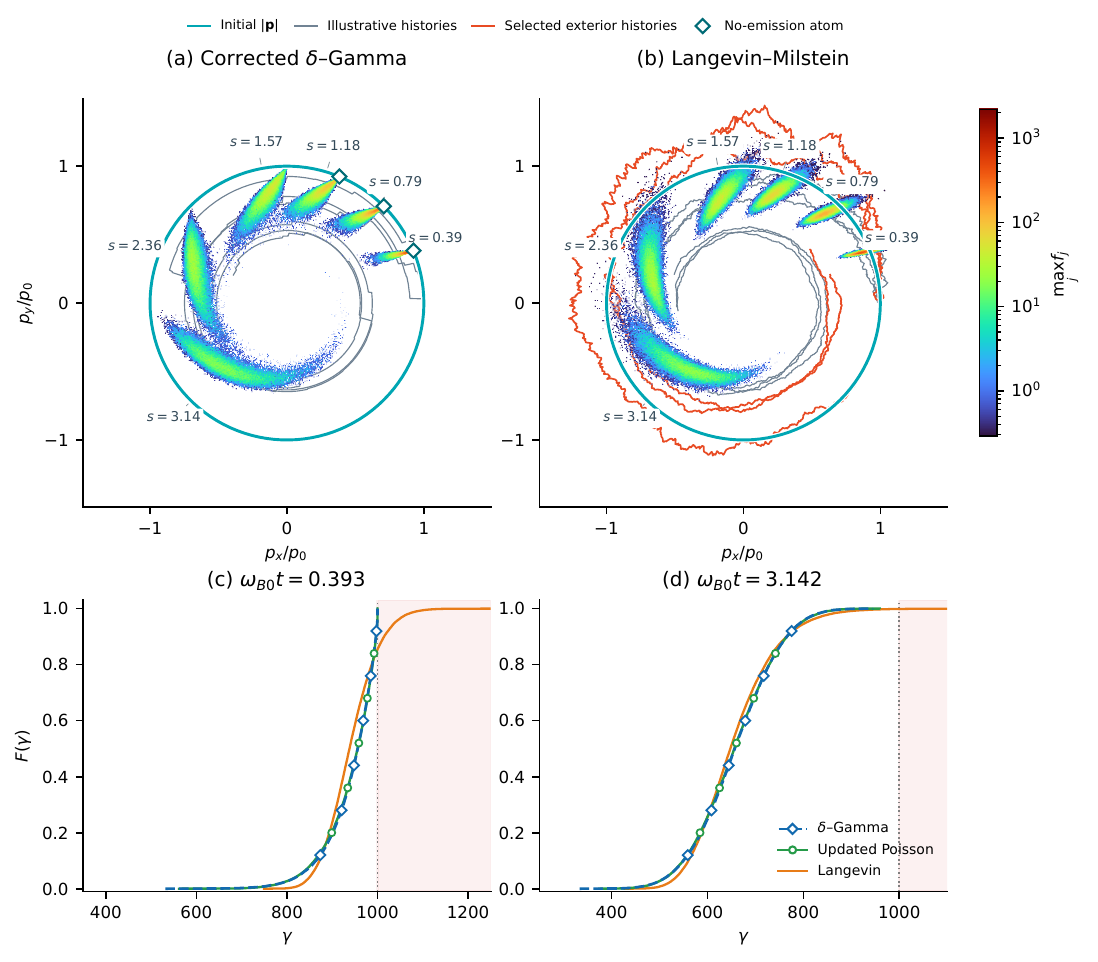}
\caption{One-sided recoil in uniform magnetic motion.
(a,b) Electron momentum densities at six indicated times for corrected
$\delta$-Gamma and Langevin-Milstein dynamics. Color gives the pointwise
maximum of separately initial-population-normalized snapshot densities,
not a time average. The turquoise circle marks the initial momentum
magnitude; diamonds on that circle identify the no-emission atom. Gray curves are
illustrative fixed-ID histories; three red histories are selected from the full
ensemble for large or sustained excursions beyond the circle. Curves span
$0\leq s\leq2\pi$; the density snapshots end at $s=\pi$. (c,d) Electron-energy CDFs, including the
state-updated explicit photon reference; shading marks $\gamma>\gamma_0$.
Here $\gamma_0=1000$, $\chi_0=0.04$,
$b=0.04/\sqrt{\gamma_0^2-1}$, and $N=4\times10^4$ per model.
Time is $s=\omega_{B0}t$. Supplement~II documents the construction
and the energy-distribution comparison under step refinement.}
\label{fig:langevin}
\end{figure*}

\begin{figure*}[t]
\centering\includegraphics[width=\textwidth]{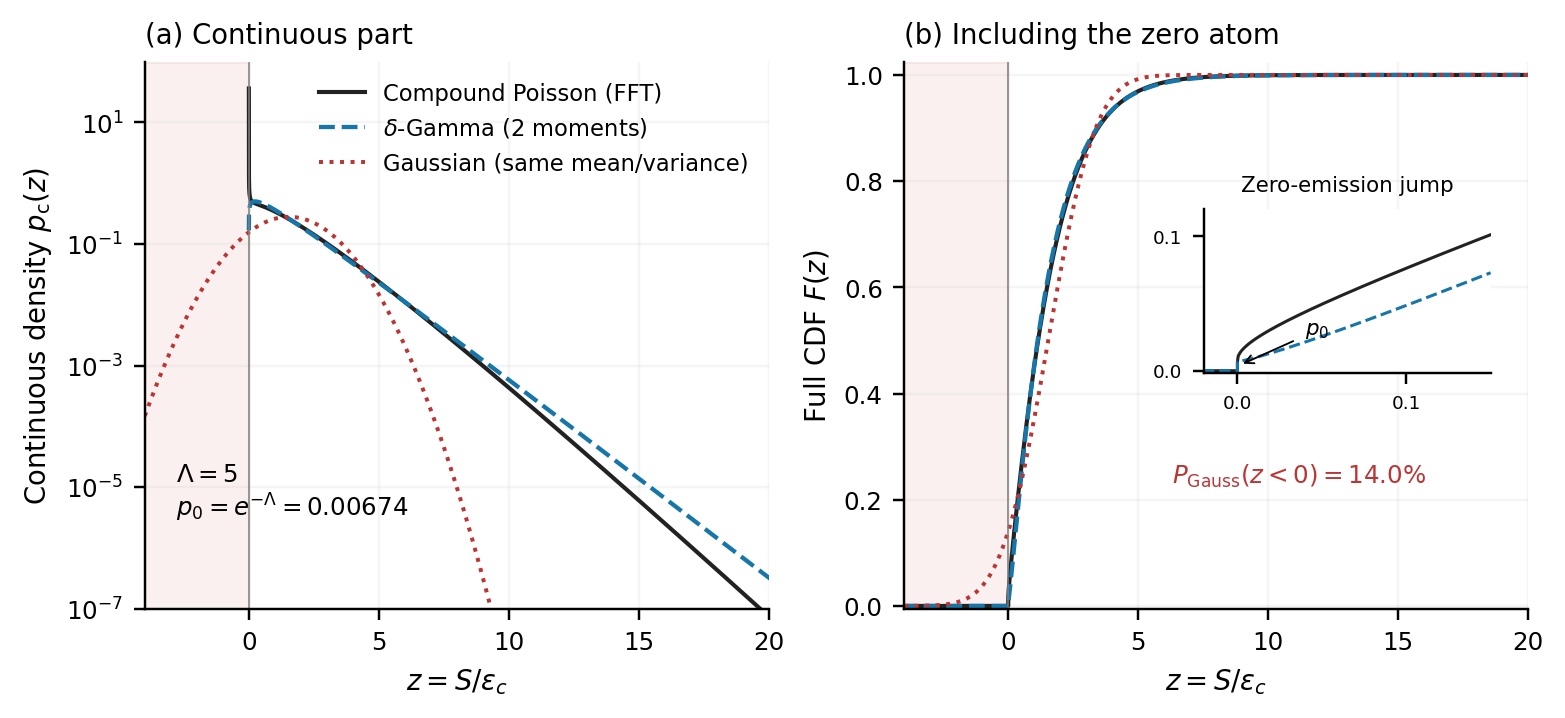}
\caption{The sign of recoil is information beyond two moments. A frozen
classical synchrotron compound-Poisson loss is compared with its conditional
Gamma closure and a Gaussian step with the same mean and variance.
The shaded region represents $S<0$, corresponding to a radiative energy gain.
The density panel shows the continuous part; the CDF also resolves the
no-emission atom. Here $\Lambda=5$ and $\epsilon_c=3\gamma\chi/2$.
The Gaussian curve represents a frozen drift--diffusion increment.
The reference construction and numerical checks are given in Supplement I.}
\label{fig:recoil_structure}
\end{figure*}

A Gaussian diffusion approximation offers a way to retain stochastic
recoil without resolving individual emissions
\cite{Neitz2013prl,Niel2018pre}. It retains the mean radiative damping
and the second recoil moment that supplies fluctuations, but its
increments have both signs and therefore admit radiative energy gains,
as illustrated by magnetic motion [Fig.~\ref{fig:langevin}].
Although useful in an appropriate limit, this approximation does not
preserve the one-sided character of emission. The underlying jump
process also has nonzero higher-order Kramers--Moyal coefficients,
which a second-order Fokker--Planck operator omits
\cite{Moyal1949,Pawula1967}.
The support difference is visible even before coupling recoil to a
trajectory: finite-step laws with identical mean and variance can
assign different probabilities to physically forbidden energy gains
[Fig.~\ref{fig:recoil_structure}].

We retain the one-sided character of energy loss by approximating its
finite-step distribution directly. The starting point is the same
local constant-crossed-field emission kernel used in photon-resolved
particle simulations
\cite{NikishovRitus1964,NarozhnyNikishovRitus1965,BaierKatkov1968,
BaierKatkov1969,Elkina2011prstab,Nerush2011prl,Duclous2011ppcf,
Ridgers2014jcp}.
A conditional Gamma closure replaces a packet of soft photons by one
positive draw while retaining its empty branch. Related positive
aggregate-loss models are established in insurance mathematics, and
Gamma processes also appear in finance
\cite{Dufresne1991,Mikosch2009,MadanCarrChang1998}.
We apply this closure along a resolved particle trajectory within
the local Markov emission model used in photon-resolved simulations.

Our construction combines positive aggregation with state feedback,
accounting for changes in the emitter without individual photon draws.
Uniform magnetic motion makes the energy boundary visible and supplies
an independent analytical cooling reference that exposes a bias shared
by frozen photon packets and their Gamma approximation. Correcting
the moments reduces this bias and improves agreement of the
electron-energy distribution with state-updated photon sampling.
Ion-channel motion connects the stochastic trajectories to directed
radiation, while focused laser scattering probes the excessive-loss
risk. Supplements~I and~II give the construction and independent
verification.

\paragraph{A positive aggregate recoil.}
We use $\hbar=c=1$, electron energies in units of $m$, and laboratory time
$\tau=t/\tc$, with $\tc=\hbar/(mc^2)$. Let $W$ be the local LCFA emission
rate per unit $\tau$ and $m_n=\langle\epsilon^n\rangle$ the raw
single-photon energy moments. Freezing the emitting state for a step $h$
gives the compound-Poisson loss
\begin{equation}
 S=\sum_{j=1}^{N}\epsilon_j,\qquad
 N\sim\operatorname{Poisson}(\Lambda),\quad\Lambda=Wh.
 \label{eq:cp}
\end{equation}
Its empty branch has weight $p_0=\ee^{-\Lambda}$; its mean and variance
are $M=hWm_1$ and $V=hWm_2$. We keep that branch and replace the
nonempty packet by
\begin{equation}
 \widetilde S=\begin{cases}
 0,&\text{with probability }p_0,\\
 \operatorname{Gamma}(k,\theta),&\text{with probability }q=1-p_0.
 \end{cases}\label{eq:mixture}
\end{equation}
Matching prescribed finite-step moments gives
\begin{equation}
\begin{aligned}
 \mu_+ &= M/q, & v_+ &= V/q-p_0M^2/q^2,\\
 k &= \mu_+^2/v_+, & \theta &= v_+/\mu_+.
\end{aligned}
\label{eq:matching}
\end{equation}
The mixture preserves the zero atom and prescribed mean and variance,
with every nonzero draw representing a loss. Its approximate shape differs
from a Gaussian step even when the two moments agree
[Fig.~\ref{fig:recoil_structure}].

One Gamma variate replaces a random number of photon draws
\cite{MarsagliaTsang2000}, reducing the expected number of positive-loss
variates by a factor $\Lambda/(1-\ee^{-\Lambda})$ at fixed state.
The opportunity grows with photon multiplicity as long as the emitter
and its kernel vary little within the packet. A useful aggregation
window therefore combines
\begin{equation}
 Wh\gtrsim1,\qquad M/(\gamma-1)\ll1,
 \label{eq:window}
\end{equation}
with a resolved orbit and controlled excessive-loss risk. Within this
window the radiation step need not resolve emissions one by one.
The reduction in photon draws is distinct from total runtime, which also
includes trajectory integration, coefficient evaluation and diagnostics.

\paragraph{Magnetic motion: preserving the energy boundary.}
The sign constraint becomes geometrical in a uniform transverse magnetic
field: because the field does no work, emission can reduce the momentum
magnitude or leave it unchanged, but cannot carry it outside the initial
circle. Corrected Gamma sampling respects this boundary, whereas the
Langevin model develops an exterior population despite using the same
first two local LCFA recoil moments [Fig.~\ref{fig:langevin}]. The energy
CDFs compare both descriptions with state-updated explicit photon sampling:
the Gamma closure closely follows the jump reference, while the early
Langevin population above the initial energy persists when its integration
step is halved. This geometrical distinction reflects the recoil statistics
in Fig.~\ref{fig:recoil_structure}: matching two moments does not preserve
the direction of energy transfer. Supplement~II gives the coefficients
and the energy-distribution refinement check. This test illustrates the
support constraint; the high-multiplicity cooling benchmark below tests
aggregation.

\paragraph{The emitter changes within a packet.}
Preserving the sign is the starting point; aggregation must also account
for changes in the emitter. Even an exact draw from Eq.~\eqref{eq:cp}
keeps the initial state throughout the radiation step and can therefore
overestimate subsequent loss as the electron cools. This bias belongs to
explicit frozen packets as well as their Gamma approximation.

\begin{figure*}[t]
\centering\includegraphics[width=\textwidth]{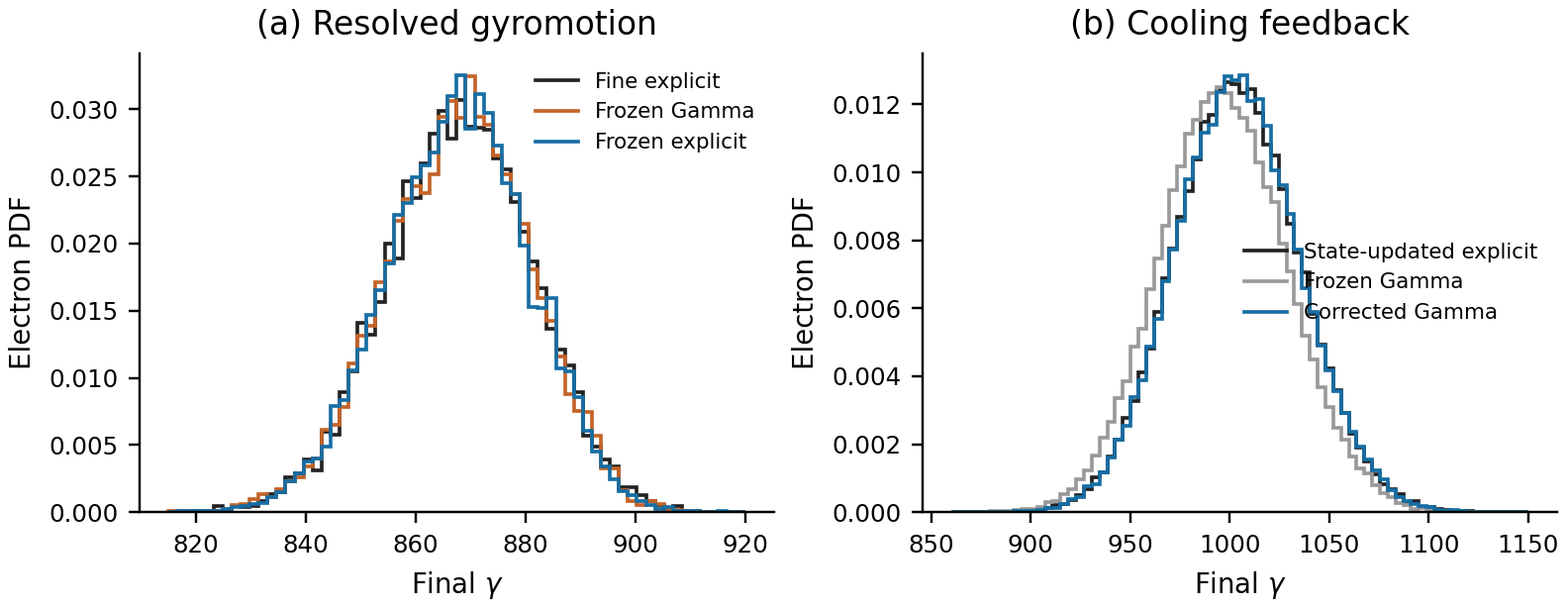}
\caption{Two magnetic checks. (a) Final electron-energy distributions during
resolved gyromotion, comparing frozen aggregate and explicit photon sampling.
(b) A separate energy-only cooling test: correcting within-step feedback removes
the main displacement of the frozen aggregate distribution relative to a
state-updated explicit reference. 
The setups and convergence scans are specified in
Supplement~II.}
\label{fig:magnetic}
\end{figure*}

Uniform magnetic motion provides an independent way to expose this
error: the analytical synchrotron moments of
Baier, Katkov and Strakhovenko \cite{BaierKatkovStrakhovenko1974}.
The resolved-orbit test first demonstrates aggregation while retaining
the electron-energy distribution [Fig.~\ref{fig:magnetic}(a)]. A separate
stronger-cooling test then reveals the common frozen-state displacement
[Fig.~\ref{fig:magnetic}(b)]. Sampler-to-sampler agreement alone would
have missed its origin.

We account for the leading feedback of recoil on subsequent emission
through the local jump generator $\mathcal L$. With $A=Wm_1$,
the mean loss becomes
\begin{equation}
 M=hA+\tfrac12h^2\mathcal LA+O(h^3).
 \label{eq:feedback}
\end{equation}
Together with the corresponding variance correction (Supplement~II),
this mean enters Eq.~\eqref{eq:matching} without changing the single-draw
update. In the stationary magnetic geometry the no-emission weight also
remains exact along the nonradiating orbit.

State feedback improves both cooling and the electron-energy distribution:
the moments agree with the independent small-$\chi$ synchrotron expansion,
while at $\chi_0=0.002$ the empirical CDF distance from the state-updated
explicit reference falls from 0.08755 to 0.00568 (Supplement~II).
This energy-only test starts with a mean load exceeding twenty photons
per aggregate step. A separate resolved-orbit scan checks trajectory
accuracy and its shared-step constraint; neither test requires the
corrected Gamma shape to be an exact transition law.

\paragraph{Aggregation window and computational opportunity.}
The aggregation capacity can be estimated from the allowed change
of the emitting state. Let $\delta_c$ be the permitted mean fractional
energy loss per step and $\delta_\phi$ the permitted magnetic rotation.
In the ultrarelativistic soft-synchrotron limit, these budgets give
\begin{equation}
 \Lambda \lesssim
 \min\left[
 \frac{15\,\delta_c}{4\sqrt{3}\,\chi},
 \frac{5\alpha\gamma}{2\sqrt{3}}\,\delta_\phi
 \right].
 \label{eq:aggregation_capacity}
\end{equation}
The cooling budget permits greater aggregation as recoil softens, whereas
the rotation budget constrains the shared trajectory/radiation step used
here. The latter is a coupling constraint, not an intrinsic limit of the
loss law: magnetic cooling is independent of rotation phase. Supplement~I
derives these estimates alongside the moment-admissibility and boundary-risk
conditions.

A separate lower-energy ion-channel control gives a dynamics-only speedup
of about 1.24 for frozen Gamma relative to frozen explicit sampling on a
common grid. This excludes spectral postprocessing and is distinct from
the corrected directed-emission example below. Supplement~I gives the
cost model and timing protocol.

\paragraph{From stochastic orbits to directed radiation.}
To connect aggregate recoil to radiation from stochastic particles, we
consider betatron motion in a prescribed linear ion channel
\cite{Kostyukov2012prstab}. Equal populations with $\gamma_0=30000$ and
$\chi_0\simeq0.011$ start at opposite transverse turning points and evolve
independently over one initial betatron period on a fixed grid.

\begin{figure*}[t]
\centering
\includegraphics[width=\textwidth]{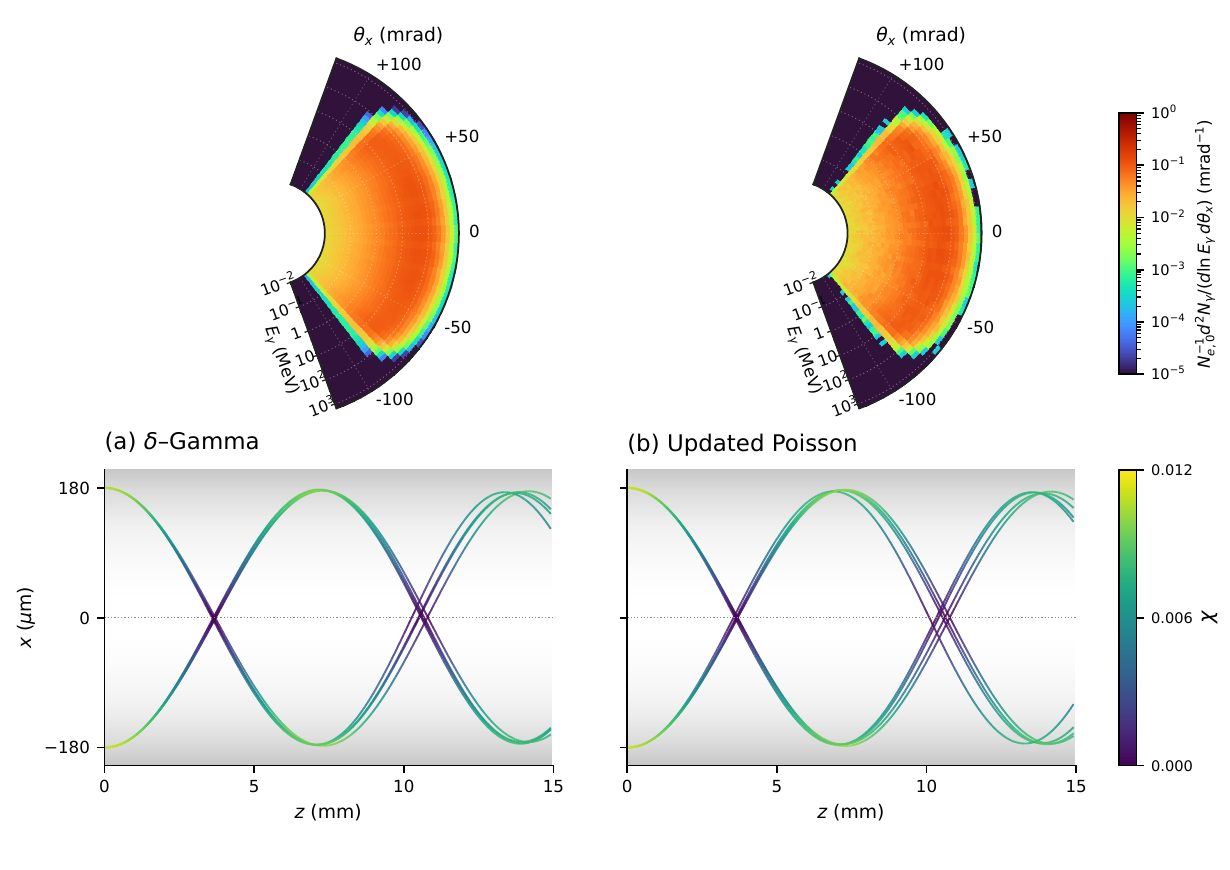}
\caption{Betatron trajectories and directed photon emission in a prescribed
ion channel. Columns compare (a) corrected $\delta$--Gamma and
(b) state-updated explicit sampling. Below: representative independent trajectories from
$x_0=\pm180\,\mu\mathrm m$, colored by $\chi$; the background indicates the
harmonic focusing potential. Above: projected photon energy--angle maps. Radius
encodes logarithmic photon energy; the expanded display angle is labeled in
the physical projected angle $\theta_x=\operatorname{atan2}(p_x,p_z)$.
Color is photon number per initial electron, per $\dd\ln E_\gamma$, and per
mrad, on a common absolute scale. The Gamma map uses a conditional mean source;
the explicit map counts accepted photons. Neither reflection averaging nor
smoothing is applied.}
\label{fig:channel_traj}
\end{figure*}

Figure~\ref{fig:channel_traj} connects the stochastic trajectories to the
projected photon energy--angle distribution. Corrected aggregate recoil
and state-updated explicit sampling retain the same main radiation
structure. The Gamma source is smoother because it averages over the
conditional photon distribution while retaining fluctuations of the
electron trajectories.

A Gamma loss does not specify its constituent photons. The diagnostic
therefore evaluates the full emission kernel along each stochastic path.
For an energy--angle bin $\mathcal A$ with local count rate
$R_{\mathcal A}$, the corrected mean source is
$hR_{\mathcal A}+h^2\mathcal LR_{\mathcal A}/2$; an analogous expression
gives radiated energy. This conditional mean-source estimator uses the same feedback order as
the recoil moments; it does not decompose individual Gamma losses into
photons. Supplement~I defines its normalization and numerical evaluation.

\begin{figure*}[t]
\centering\includegraphics[width=\textwidth]{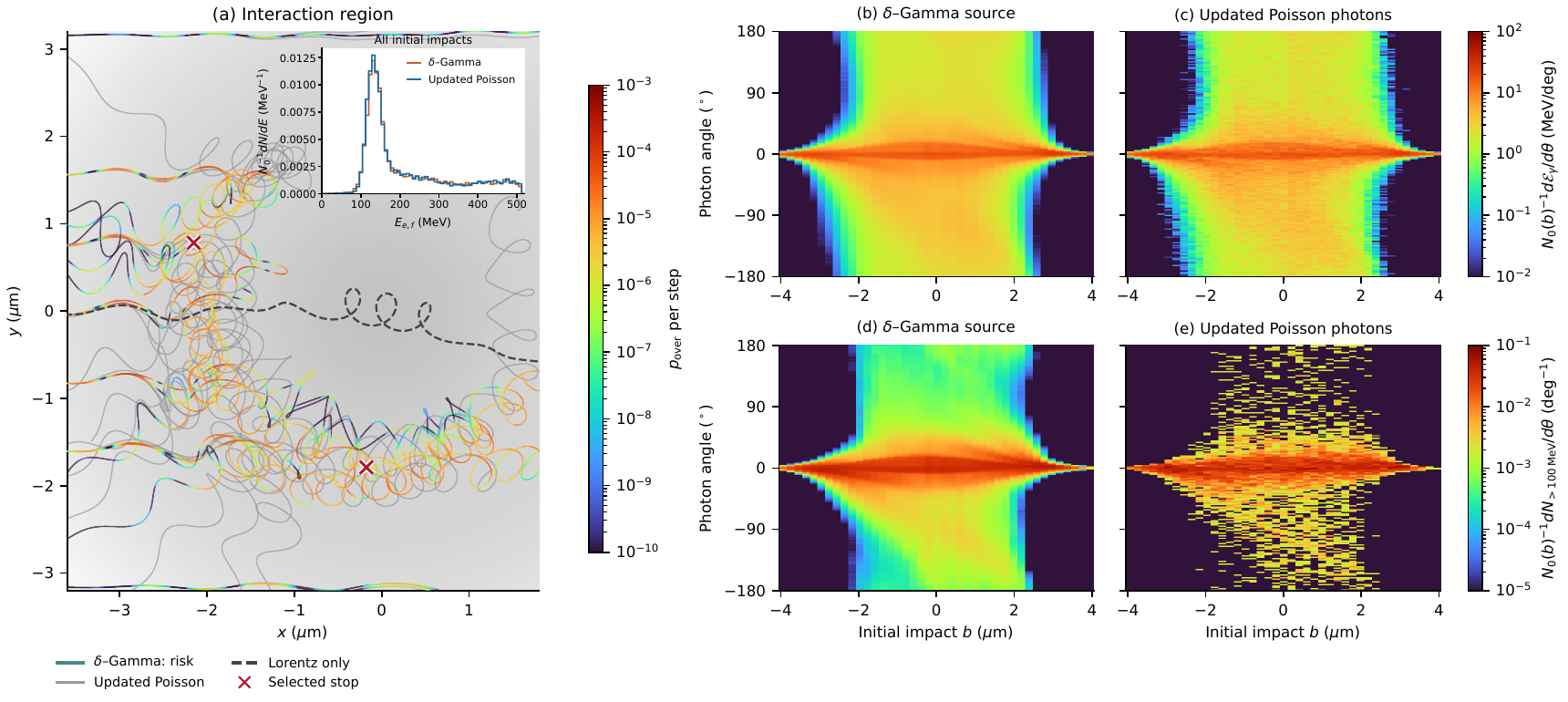}
\caption{Focused scattering and its local aggregation risk.
(a) Selected trajectories in the interaction region. Corrected Gamma paths
are colored by their excessive-loss probability before sampling; gray paths
use state-updated explicit emission. The central Lorentz-only path has no
radiation reaction. Crosses mark two explicitly selected failed histories.
Inset: final electron-energy distributions pooled over all equally populated
initial impacts and normalized to the initial ensemble; stopped histories
leave missing endpoint mass. (b,c) Radiated energy and (d,e) photon yield above
100 MeV versus initial impact and photon angle from the incoming axis.
Gamma panels evaluate a conditional mean source; explicit panels count
accepted photons. Each pair shares an absolute scale, normalized per initial
electron at that impact and per degree. There is no beam or impact-width
weighting. Gamma histories contribute until stopping and are not replaced.
Field parameters and failure accounting are specified in Supplement~I.}
\label{fig:focused}
\end{figure*}

\paragraph{Focused scattering: locating the limit.}
Focused laser scattering deliberately pushes the model towards the edge
of soft-recoil aggregation. We use counterpropagating beams following
Ref.~\cite{Elkina2014}, with combined $a_0=1000$ and initial
$\gamma_0=1000$. Figure~\ref{fig:focused} combines impact--angle maps of the recorded
radiation with trajectories that locate rare excessive losses, testing
the closure and its risk diagnostic near the limits of soft-recoil aggregation.

A positive Gamma loss cannot produce a radiative gain, but its unbounded
tail can exceed the available kinetic energy. This risk is known before
the draw:
\begin{equation}
 p_{\Gamma,\mathrm{over}}
 =q\,Q\!\left(k,\frac{\gamma-1}{\theta}\right),
 \label{eq:risk}
\end{equation}
where $Q$ is the regularized upper incomplete Gamma function. Coloring
trajectories by this local probability locates the vulnerable parts of
the interaction. The crosses show two deliberately selected stopped
histories, not a frequency sample. The radiation maps retain contributions
until stopping and normalize to the initial ensemble; the missing
continuation is not replaced.

The explicit reference updates the emitter after each accepted photon,
separating the Gamma-tail limit from the energy-budget error of frozen
photon packets. In the signed-impact ensemble, accumulated conditional
Gamma risk is 93.53 and 93 histories stop (Supplement~I). The risk therefore
provides a quantitative diagnostic, as well as a possible switching
criterion. Correcting the moments reduces cooling bias but does not
truncate the Gamma tail; step refinement alone need not remove it.
Supplement~I gives the small-step limit, reference algorithm and hybrid
controls.

\paragraph{Discussion.}
Several soft emissions can be represented by a single positive stochastic
recoil while retaining the no-emission branch and the selected finite-step
loss moments. The radiation update can therefore aggregate photons on a
resolved particle trajectory rather than follow their individual emissions.
The demonstrated window includes an initial mean load exceeding twenty
photons per aggregate step. Its extent is set by state variation and
excessive-loss risk and, in this implementation, by resolution of the
shared trajectory step, rather than by a requirement of fewer than one
photon per step.

Aggregation also exposes an error distinct from the distributional closure.
Independent analytical synchrotron moments identify a cooling bias shared
by frozen photon packets and their Gamma approximation. State feedback
substantially reduces this bias without resolving individual emissions,
and the corrected energy distribution approaches the state-updated explicit
reference. Thus reducing sampling work need not entail retaining the
leading frozen-state error.

Positive support is essential when the direction of energy transfer matters.
Matching the first two recoil moments in a Fokker--Planck--Langevin model
does not preserve that support: the magnetic comparison exhibits radiative
energy gains that persist under integration-step refinement. A diffusion
approximation can still describe observables insensitive to this difference;
the positive aggregate construction retains the energy boundary without
requiring such insensitivity.

The computational opportunity grows as individual recoils soften: the
mean-cooling budget permits photon loads scaling as $\chi^{-1}$, subject
to kernel variation, boundary risk and the chosen coupling to particle
motion. The realized speedup depends on photon multiplicity and the costs
of feedback, trajectory integration and diagnostics. Directed betatron emission shows
how aggregate trajectories can support kernel-based radiation diagnostics,
while focused scattering supplies a computable warning of excessive-loss
risk. Together these results establish a practical route to stochastic
soft-photon recoil in packets, with explicit controls on where that
reduction remains useful.
\begin{acknowledgments}
This work was supported by the Deutsche Forschungsgemeinschaft
(DFG) and the BMBFTR.
\end{acknowledgments}
\bibliographystyle{apsrev4-2}
\bibliography{references}
\end{document}


\title{
Supplement I: Construction, implementation, and field tests
}
\author{Nina Elkina$^1$ and Alexander Pukhov$^{1}$}
\affiliation{$^1$Institut f\"ur Theoretische Physik I, Heinrich-Heine-Universit\"at D\"usseldorf, 40225 D\"usseldorf, Germany}
\date{November 19, 2026}
\maketitle

\section{Scope and units}
This supplement defines the positive aggregate update and its coupling
to particle motion. The corrected Gamma sampler is the principal method;
state-updated explicit photon sampling supplies its local reference.
Frozen packets, hybrid switches and alternative closures are controls
used to identify particular errors. The sections below first specify the
construction, reference and diagnostics, then document their numerical
implementation and field tests. Supplement~II provides the independent
synchrotron cooling analysis and magnetic verification. The
construction is a closure of this common local Markov emission model;
its accuracy is assessed separately from the approximations entering
that model. Aggregating sequential emissions does not supply genuine
multiphoton QED probabilities.

We adopt $\hbar=c=1$, energies in units of the electron mass $m$, momenta
$\mathbf p/(mc)$, and laboratory time $\tau=mt=t/\tc$,
$\tc=\hbar/(mc^2)$. Dimensionless fields are $|e|\mathbf E/m^2$ and
$|e|\mathbf B/m^2$ in natural units. A rate per unit $\tau$ becomes a rate
per laboratory second upon division by $\tc$. A dimensionless photon energy
$\epsilon$ corresponds to $mc^2\epsilon$ in physical units.
For these field and momentum variables,
\begin{equation}
 \chi=\sqrt{(\gamma\mathbf E+\mathbf p\times\mathbf B)^2
                   -(\mathbf p\cdot\mathbf E)^2}.
\end{equation}
Thus a transverse magnetic field gives $\chi=b\sqrt{\gamma^2-1}$,
$b=B/B_{\rm cr}$, with $B_{\rm cr}=m^2/|e|$ in these units.

\section{Common emission kernel}
Let $s=\epsilon/\gamma$ be the photon energy fraction. The spin-averaged
constant-crossed-field rate per unit natural laboratory time is
\begin{align}
 \frac{\dd W}{\dd s}&=\frac{\alpha}{\sqrt3\pi\gamma}
 \left[\left(1-s+\frac1{1-s}\right)K_{2/3}(z)
                -\int_z^\infty K_{1/3}(y)\,\dd y\right],\\
 z&=\frac{2s}{3\chi(1-s)},\qquad 0<s<1-\frac1\gamma.
 \label{eq:kernel}
\end{align}
Equivalently, the bracket is
$\int_z^\infty K_{5/3}(y)\,\dd y+s^2K_{2/3}(z)/(1-s)$
\cite{Ritus1985,DiPiazza2012rmp}.
The rate, CDF and raw moments use the same kinetic-energy cutoff:
\begin{equation}
 R_n=\int(\gamma s)^n\frac{\dd W}{\dd s}\,\dd s,
 \qquad W=R_0,\qquad m_n=\frac{R_n}{W}.
 \label{eq:moments}
\end{equation}
The differential rate per photon energy is
$w(x,\epsilon)=\gamma(x)^{-1}(\dd W/\dd s)|_{s=\epsilon/\gamma(x)}$.
The $m_n$ are raw single-photon moments. A subscript $+$ below denotes
conditioning on a nonempty aggregate step. The fractional single-photon
moments are $\mu_n=m_n/\gamma^n$. We write $F(\epsilon\mid x)$ for the
normalized photon-energy CDF at emitting state $x$, with $F=0$ below zero
and $F=1$ above the available kinetic energy.

All comparisons adopt the ultrarelativistic, collinear LCFA kernel in
Eq.~\eqref{eq:kernel}. This common kinetic model fixes the emission physics
against which the aggregate closure is tested.

The numerical kernel table covers $10^{-10}\leq\chi\leq30$ with 192 logarithmic
kernel nodes and 6000 transformed-energy nodes over $10^{-18}\leq z\leq120$.
The sampled CDF is interpolated linearly in $\log\chi$ and the rate
logarithmically. Moments are evaluated from that sampled CDF using two-point
Gaussian quadrature in each $z$ interval, with an auxiliary 765-point
$\chi$ grid. The finite-energy cutoff is handled directly when its
transformed location is below $z=40$. Supplement II additionally checks the
table against direct Bessel-function quadrature in the tested magnetic range.

\section{Conditional Gamma construction}
For a frozen interval $h$, let $N\sim\operatorname{Poisson}(\Lambda)$,
$\Lambda=Wh$, and $S=\sum_{j=1}^N\epsilon_j$. Independence gives
\begin{equation}
 \log\mathbb E[\ee^{iuS}]
 =\Lambda\{\mathbb E[\ee^{iu\epsilon}]-1\}.
\end{equation}
Consequently its cumulants are
\begin{equation}
 \kappa_n(S)=\Lambda m_n=hR_n,
 \label{eq:cp_cumulants}
\end{equation}
with raw single-photon moments on the right-hand side. In particular,
\begin{equation}
 \mathbb E S=\Lambda m_1,\quad
 \mathbb E S^2=\Lambda m_2+\Lambda^2m_1^2,\quad
 \operatorname{Var}S=\Lambda m_2.
\end{equation}
Since every positive raw loss moment vanishes on the zero branch of
weight $p_0=\ee^{-\Lambda}$, conditioning on its complement $q=1-p_0$ gives
\begin{equation}
 \mu_+=\frac{\Lambda m_1}{q},\qquad
 v_+=\frac{\Lambda m_2}{q}-\frac{p_0\Lambda^2m_1^2}{q^2}.
 \label{eq:conditional}
\end{equation}
A Gamma variate with shape $k=\mu_+^2/v_+$ and scale
$\theta=v_+/\mu_+$ has these conditional moments. Recombining the branches
therefore recovers the exact zero mass, mean and variance of the frozen model.
This algebraic identity holds for the input moments supplied by the
numerical tables.

Positivity of $v_+$ follows directly from
\begin{equation}
 v_+=\frac\Lambda q(m_2-m_1^2)
       +\frac{\Lambda m_1^2}{q^2}(q-p_0\Lambda).
\end{equation}
Here $m_2\geq m_1^2$ and
$q-p_0\Lambda=\ee^{-\Lambda}(\ee^\Lambda-1-\Lambda)\geq0$.
The zero-variance limit is a deterministic positive loss.

For arbitrary prescribed finite-step mean $M$ and variance $V$, the same
conditioning identity yields
\begin{equation}
 \mu_+=M/q,\qquad v_+=V/q-p_0M^2/q^2.
 \label{eq:general_matching}
\end{equation}
The same identity applies to the corrected finite-step moments below.
For a nonzero rate, $M$ and $V$ must be finite and positive, and the
conditional variance must satisfy $v_+>0$. These are admissibility conditions
on the moment closure, distinct from its excessive-loss risk.

\subsection{Stable evaluation and limiting behavior}
We compute $q=-\operatorname{expm1}(-\Lambda)$ and skip zero-rate updates.
For small $\Lambda$, write $r=\Lambda/q$ and evaluate
\begin{equation}
 v_+=r\{m_2+m_1^2(\Lambda-r)\},\qquad
 \Lambda-r=-1+\frac\Lambda2-\frac{\Lambda^2}{12}
                           +\frac{\Lambda^4}{720}+\cdots .
\end{equation}
This avoids cancellation of small probabilities.
As $\Lambda\to0$, the positive branch tends to a moment-matched single-photon
Gamma law. Its finite-step shape is specified by the Gamma closure. Parameters
are refitted with the state and step, retaining the zero mass and the selected
moments at each update.

\subsection{Auxiliary shape diagnostic}
The main update matches two moments. An optional third-moment diagnostic
assesses the remaining shape error by comparing
\begin{align}
 r_{2,+}&=(\Lambda m_2+\Lambda^2m_1^2)/q,\\
 r_{3,+}&=(\Lambda m_3+3\Lambda^2m_1m_2+\Lambda^3m_1^3)/q,\\
 \mu_{3,+}^{\rm CP}&=r_{3,+}-3\mu_+r_{2,+}+2\mu_+^3,\qquad
 \mu_{3,+}^{\Gamma}=2v_+^2/\mu_+.
\end{align}
The reported hybrid uses the boundary-risk criterion below; the third-moment
mismatch supplies a separate shape diagnostic.

\section{Fixed-step coupling to particle motion}
A trajectory step contains an implicit-midpoint Lorentz half-step, a radiation
update of duration $h$, and a second midpoint half-step. Kernel coefficients
are evaluated at the intermediate state. During the radiation stage its
position and local external field are held fixed. All radiation updates use this fixed
grid. The explicit frozen reference draws $N$ and then $N$ photons from the
same intermediate-state CDF, followed by one aggregate recoil update.

For $S<\gamma-1$ the common collinear map is
\begin{equation}
 \gamma'=\gamma-S,\qquad
 \mathbf p'=\sqrt{(\gamma')^2-1}\,\frac{\mathbf p}{|\mathbf p|}.
\end{equation}
Momentum here is in units of $mc$. Radiation reaction enters through the
sampled recoil alone. The Lorentz integrator evolves three-momentum, and the
radiation map sets the recoiled momentum on the mass shell. Midpoint rotation details are given
in Supplement II and are related to the numerical considerations of
Ref.~\cite{Elkina2014}.

\section{State feedback and admissibility}
\label{sec:feedback}
Let $x$ denote the momentum state at the radiation midpoint, with position
and external field fixed, and let $T_\epsilon x$ be its collinear recoil map.
The local one-photon generator is
\begin{equation}
 (\mathcal Lf)(x)=\int_0^{\gamma(x)-1}
 [f(T_\epsilon x)-f(x)]w(x,\epsilon)\,\dd\epsilon,
 \label{eq:local_generator}
\end{equation}
where $w$ is the differential emission rate per unit photon energy and
natural laboratory time. Define $A(x)=R_1(x)$ and $B(x)=R_2(x)$.
For the cumulative energy loss, the generator expansion gives
\begin{align}
 M_h&=hA+\tfrac12h^2\mathcal LA+O(h^3),\\
 V_h&=hB+h^2(\tfrac12\mathcal LB+C)+O(h^3),\\
 C(x)&=\int\epsilon[A(T_\epsilon x)-A(x)]w(x,\epsilon)\,\dd\epsilon.
 \label{eq:feedback_moments}
\end{align}
The numerical closure inserts the displayed terms through $h^2$ into
Eq.~\eqref{eq:general_matching}. The derivation and independent magnetic
checks are given in Supplement II. At each quadrature recoil the code
recomputes the post-recoil energy, mass-shell momentum, $\chi$, rate and
cutoff in the same fixed field; the dependence $\chi\propto\gamma$ is used
only when the local scaling check permits it.

The no-event weight remains $p_0=\exp[-W(x)h]$. This is exact for the local
radiation stage even when accepted emissions subsequently change the rate:
until the first event the state is unchanged. In an evolving external field,
the full nonradiating orbit can have a time-dependent rate. Its no-event
weight involves the integral of that rate; the fixed-midpoint stage
approximates it as part of the trajectory splitting.

\subsection{Principal update, reference, and diagnostic controls}
Corrected Gamma is the proposed feedback update. Frozen Gamma supplies its
baseline; the state-updated explicit process is the main local reference.
The remaining variants diagnose frozen-packet error or risk switching:

\begin{center}
\begin{tabular}{lp{95mm}}\toprule
Sampler & Radiation update\\\midrule
Frozen Gamma & The zero atom and two moments of the frozen compound-Poisson law.\\
Corrected Gamma & The same zero atom, with the two feedback-corrected moments.\\
Frozen explicit & A Poisson photon count, all energies drawn at the initial radiation state.\\
State-updated explicit & Each accepted photon changes the emitting state before the next proposal.\\
Hybrid & Gamma, with a frozen explicit fallback selected by boundary risk.\\
Corrected hybrid & Corrected Gamma, with a frozen explicit fallback selected by boundary risk or inadmissible corrected moments.\\\bottomrule
\end{tabular}
\end{center}
The hybrid variants retain a frozen explicit fallback and are controls of
risk-based switching, not guarantees of completion. For inadmissible
corrected moments, the corrected Gamma update stops before
sampling; the corrected hybrid instead uses the explicit frozen fallback.
The implementation records these cases separately from excessive-loss stops.
Neither changes the fixed external step or redraws an excessive recoil.
The correction matches the selected finite-step moments; it does not make
the Gamma transition law exact or establish a global second-order sampler.

\section{State-updated explicit reference on a fixed grid}
The frozen compound-Poisson control is complemented by a local state-dependent
one-photon jump process. During the radiation stage the midpoint field and
position remain fixed. After each accepted photon we update the electron's
mass-shell momentum, energy, $\chi$, rate and kinetic-energy cutoff.
For a bound $\Omega\ge W(x)$ on every accessible recoil state, define
$P=I+\mathcal L/\Omega$. Uniformization gives
\begin{equation}
 \exp(h\mathcal L)=\ee^{-\Omega h}
 \sum_{n=0}^{\infty}\frac{(\Omega h)^n}{n!}P^n.
\end{equation}
We draw $n\sim\mathrm{Poisson}(\Omega h)$ proposals. At state $x$, each proposal
is accepted with probability $W(x)/\Omega$; an accepted photon is sampled from
the current normalized kernel. A rejected proposal leaves the state unchanged.
These null proposals are not redraws of inadmissible photons. Event times and
radiation substeps are not generated. The accepted photon count generally is
not Poisson with parameter $W(x_0)h$.

For the implemented CCF table, let $I(\chi)$ denote its uncut rate integral.
The laboratory-time rate is $\alpha I(\chi)/(\sqrt3\pi\gamma)$, and
$\chi/\gamma\le |\mathbf E|+|\mathbf B|$. Thus a numerical majorant is
\begin{equation}
 \Omega=\frac{5\alpha}{2\sqrt3}\rho
 (|\mathbf E|+|\mathbf B|),\qquad
 \rho=\max\left(1,\max_i\frac{2I_i}{5\pi\chi_i}\right)(1+10^{-12}).
\end{equation}
Log--log interpolation places the maximum of $I(\chi)/\chi$ at a table node;
the low-$\chi$ continuation has normalized ratio one. The cutoff only reduces
the rate. The code checks the majorant at every proposal and the photon energy
ledger after every radiation stage.

This construction samples the local semigroup of the numerical kernel. Field
freezing, collinear recoil and outer Lorentz splitting remain approximations.
Its local no-emission weight is the one given in
Sec.~\ref{sec:feedback}. Completion and failure statistics for the two focused
ensembles and the lower-energy channel control are reported with their setups
below. Their explicit spectra count accepted photons.

\section{Available-energy boundary and risk accounting}
Positive support excludes radiative energy gains, but a Gamma law has no
finite upper bound. The predicted excessive-loss probability is
\begin{equation}
 p_{\Gamma,\rm over}
 =q\,Q\!\left(k,\frac{\gamma-1}{\theta}\right).
\end{equation}
This expression evaluates the boundary risk of the Gamma law before a draw.
The focused hybrid switches to explicit frozen photon sampling when the risk
exceeds its threshold. The two contributions to boundary failure are assessed
separately: the Gamma tail and excessive sums in the frozen reference.
For $S\geq\gamma-1$ a trajectory is recorded and stopped. Losses are never
clipped, rejected and redrawn, or replaced by another trajectory.

Let $\mathcal F_j$ contain the state and all history before the $j$th
radiation draw. Let $I_j$ indicate an active Gamma update, after any hybrid
switch, and let $p_j$ be its excessive-loss probability. Both are determined
by $\mathcal F_j$. If $Y_j$ indicates failure of that Gamma draw, then
$\mathbb E[I_jY_j\mid\mathcal F_j]=I_jp_j$, and hence
\begin{equation}
 \mathbb E\sum_j I_jY_j
 =\mathbb E\sum_j I_jp_j.
 \label{eq:risk_compensator}
\end{equation}
Thus $H=\sum_jI_jp_j$ is an accumulated conditional-risk diagnostic. Its
realized value is not the failure probability of that realized path.
With stopping at the first boundary failure, the number of Gamma failures on
a path is zero or one; its ensemble expectation is the probability of stopping
through the Gamma branch. Explicit-Poisson failures are a separate contribution.
A fixed per-step threshold does not fix $H$ when the number of steps changes.

\subsection{Small-step limit of excessive-loss risk}
Reducing the radiation step controls within-step state variation but does
not recover the exact single-photon spectrum from a two-moment Gamma fit.
At a fixed emitting state with $W>0$ and $m_2>m_1^2$, both the frozen and
feedback-corrected constructions have the limits
\begin{equation}
 k_h\longrightarrow k_1=\frac{m_1^2}{m_2-m_1^2},\qquad
 \theta_h\longrightarrow\theta_1=\frac{m_2-m_1^2}{m_1}.
\end{equation}
Writing $r_\Gamma(x)=Q(k_1,(\gamma-1)/\theta_1)$ gives
\begin{equation}
 p_{\Gamma,\rm over}(h\mid x)=hW(x)r_\Gamma(x)+O(h^2).
 \label{eq:small_step_risk}
\end{equation}
The risk per attempt decreases with $h$, while the number of attempts over
a fixed duration increases. Thus the local excessive-loss rate can remain
nonzero as $h\to0$. For a prescribed smooth sequence of emitting states,
the leading accumulated conditional risk is the integral of
$W(x)r_\Gamma(x)$ over the active intervals. For stochastic, stopped histories,
the conditional-risk accounting in Eq.~\eqref{eq:risk_compensator} applies;
a realized risk sum is not a pathwise failure probability.

This residual shape error is distinct from frozen-state cooling bias:
a smaller step can improve the cooling moments without eliminating
excessive-loss stops. Their counts need not vary monotonically because
$r_\Gamma$ depends on the local state, not on a universal threshold
in $\chi$.

The aggregation window combines appreciable $Wh$, small fractional cooling,
weak local-kernel variation, orbit resolution and acceptable accumulated risk.
Neither small mean cooling nor a universal $\chi$ threshold controls every
rare loss. Boundary risk, shape error and frozen-state error are different
quantities.
\subsection{Estimates of aggregation capacity}

Let $\Lambda=Wh$ denote the local frozen photon load.
Requiring the mean loss to satisfy
$M/(\gamma-1)\lesssim\delta_c$, with $M=\Lambda m_1$,
gives
\begin{equation}
 \Lambda_{\mathrm{cool}}
 =\frac{\delta_c(\gamma-1)}{m_1}.
\end{equation}
Using the leading soft-synchrotron moment
\begin{equation}
 \frac{m_1}{\gamma}
 =\frac{4\sqrt{3}}{15}\chi+O(\chi^2)
\end{equation}
and $\gamma\gg1$ yields
\begin{equation}
 \Lambda_{\mathrm{cool}}
 \simeq\frac{15}{4\sqrt{3}}\frac{\delta_c}{\chi}.
\end{equation}

For transverse motion in a uniform magnetic field,
the local rotation during a step is
$\Delta\phi\simeq bh/\gamma$, where $h$ is measured
in units of $\tau_{\mathrm C}$.
Combining $\Delta\phi\lesssim\delta_\phi$ with
$W\simeq5\alpha b/(2\sqrt{3})$ gives
\begin{equation}
 \Lambda_{\mathrm{orbit}}
 \simeq\frac{5\alpha\gamma}{2\sqrt{3}}\delta_\phi.
\end{equation}
The cooling estimate scales as $\chi^{-1}$, whereas
the rotation estimate scales as $\gamma$ and is
independent of $b$ at this order. The rotation budget constrains the
shared trajectory/radiation step used here, not the aggregate loss law
itself. In a uniform magnetic field the energy-loss rate is independent
of rotation phase; this budget controls the resolved orbit, not a
phase-induced cooling error.

For the corrected sampler, the fractional loss is evaluated
using its corrected mean $M_h$. The expressions above provide
leading estimates, rather than bounds on the remaining
moment-closure error. Both budgets should be evaluated along the evolving
trajectory, together with moment admissibility and accumulated
excessive-loss risk. They estimate aggregation capacity; they do not
constitute an additional acceptance rule used in the reported fixed-grid runs.

\begin{table}[tbp]
\centering
\caption{Leading aggregation estimates for a mean-loss budget
$\delta_c=0.02$ and a rotation budget $\delta_\phi=0.1$ rad.
The last column is the smaller of the two estimates,
before applying moment-admissibility and boundary-risk checks.}
\begin{tabular}{rrrrr}
\toprule
$\gamma$ & $\chi$ &
$\Lambda_{\mathrm{cool}}$ &
$\Lambda_{\mathrm{orbit}}$ &
$\min(\Lambda_{\mathrm{cool}},\Lambda_{\mathrm{orbit}})$\\
\midrule
2000  & 0.002 & 21.7  & 2.11 & 2.11\\
20000 & 0.002 & 21.7  & 21.1 & 21.1\\
20000 & 0.001 & 43.3  & 21.1 & 21.1\\
\bottomrule
\end{tabular}
\end{table}

\section{Photon-source estimator}
\label{sec:source}
An aggregate loss does not specify constituent photon energies. The
radiation diagnostic evaluates the original kernel along the stochastic
electron trajectories. For a photon-energy bin $\mathcal A=[E_a,E_b)$,
define the count and energy rates
\begin{equation}
 R_{\mathcal A}(x)=\int_{\mathcal A}w(x,\epsilon)\,\dd\epsilon,
 \qquad J_{\mathcal A}(x)=\int_{\mathcal A}\epsilon w(x,\epsilon)\,\dd\epsilon.
\end{equation}
For the frozen stage, the conditional mean count is
$hR_{\mathcal A}=\Lambda[F(E_b\mid x)-F(E_a\mid x)]$.
Using the generator in Eq.~\eqref{eq:local_generator}, the corrected sources are
\begin{align}
 \Delta\overline N_{\mathcal A}&=hR_{\mathcal A}
                 +\tfrac12h^2\mathcal LR_{\mathcal A},\\
 \Delta\overline{\mathcal E}_{\mathcal A}&=hJ_{\mathcal A}
                 +\tfrac12h^2\mathcal LJ_{\mathcal A}.
 \label{eq:source}
\end{align}
An all-energy bin has $J_{\mathcal A}=A$, so at a fixed starting state the
mean-source energy and prescribed mean recoil have the same generator
expansion. This is a local expectation identity, not equality between the
realized loss and diagnostic energy on each path. Finite energy coverage,
quadrature and stopping must also be accounted for in recorded spectra.

The source is evaluated only on completed radiation updates. For a stopped
history its previously accumulated source is retained, and no contribution
is supplied for the failed update or its uncomputed continuation. Endpoint
and source histograms remain normalized to the initial ensemble. Thus the
local matching identity is not a claim of energy equality for censored
ensembles.

The corrected sources use 12 recoil-quadrature nodes. Bin count rates follow
from the kernel CDF; bin energy rates use two-point quantile quadrature.
Recoil integration recomputes the emitter in the same fixed local field.
Source-admissibility checks flag nonfinite values or negative contributions
beyond numerical tolerance, separately from trajectory stops. The selected source checks change populated-bin counts by at most
$2.6\times10^{-4}$ under recoil-quadrature refinement. The binned energy
source agrees with the corrected mean to about $3.1\times10^{-4}$; finer
energy bins reduce that discrepancy to about $2.6\times10^{-5}$.

The focused diagnostic evaluates only the required observables: total photon
number, total radiated energy and number above 100 MeV. Its count sources agree
with the common binned operator, and the selected hard-source test changes by
at most $3.4\times10^{-7}$ when refining from 12 to 48 recoil nodes.
The total energy source uses the corrected mean directly.

Angular emission is collinear with the electron momentum at the radiation
midpoint. A projected angle density is not a density per solid angle.
The aggregate maps average over the conditional photon distribution while
retaining stochastic variation of the electron paths. The same estimator
could be applied to explicit trajectories; its reduced photon-sampling noise
does not by itself establish improved accuracy in a rare tail. Where quoted, binwise uncertainties are evaluated over electrons, retaining
correlations among emissions from a single history.

\section{Numerical evaluation of feedback and source terms}
The header-only C++17 implementation integrates three-momentum in natural
laboratory time with a generic field callback. The corrected sampler
uses Eqs.~\eqref{eq:local_generator}--\eqref{eq:feedback_moments}, while the
Baier--Katkov--Strakhovenko result remains its independent magnetic reference.
For a recoil $\epsilon$ it recomputes $\gamma-\epsilon$ and the collinear
post-recoil momentum to obtain the local post-recoil $\chi$.

An ultrarelativistic lookup stores the three generator coefficients on 1025
logarithmic $\chi$ nodes, using 256-point recoil quadrature. When a local
geometry/cutoff check rejects $\chi\propto\gamma$ scaling, the coefficients are
computed by direct 32-point local-field quadrature. Selected-field comparisons
against 512-point quadrature give relative errors up to $1.9\times10^{-6}$ in
the finite-step mean and $2.4\times10^{-5}$ in the finite-step variance at the
focused benchmark step. The selected generator coefficients themselves have
larger relative errors, up to 1.64\%, where their finite-step contribution is
small. These checks specify the numerical closure used in the field runs.

\section{Computational gain from aggregation}
At fixed state, explicit frozen sampling draws a Poisson count and an expected
$\Lambda$ individual photon energies. The aggregate sampler performs a branch
decision and draws a Gamma variate with probability $q=1-\exp(-\Lambda)$.
The photon-variate to Gamma-variate ratio is therefore
\begin{equation}
 \mathcal R_{\rm draw}=\frac\Lambda q
 =\mathbb E[N\mid N\geq1].
\end{equation}
It approaches one in the single-photon limit and grows as $\Lambda$ for large
multiplicity. For an evolving sequence of specified states, the corresponding
expected draw-count ratio is $\sum_j\Lambda_j/\sum_jq_j$.
This reduction concerns positive-loss variates; branch decisions, count
sampling, coefficient evaluation, and orbit integration have their own costs.

Let $C_0$ denote the common trajectory and coefficient work in one step,
$C_N$ Poisson-count sampling, $C_\epsilon$ a photon-energy draw, $C_b$ a
Gamma-branch decision, and $C_\Gamma$ a Gamma draw. Then
\begin{equation}
 C_{\rm P}=C_0+C_N+\Lambda C_\epsilon,\qquad
 C_{\rm G}=C_0+C_b+qC_\Gamma+C_{\rm corr}.
\end{equation}
Here $C_{\rm corr}$ accounts for feedback coefficients when that variant is
used and is zero for the baseline frozen closure. These expressions separate the sampling savings from common trajectory
work and from any additional full-kernel spectral diagnostic.

Aggregation also relaxes an emission-based step restriction. A zero-or-one
approximation omits probability
\begin{equation}
 \Pr(N\geq2)=1-\ee^{-\Lambda}(1+\Lambda)
             =\tfrac12\Lambda^2+O(\Lambda^3).
\end{equation}
A small prescribed omitted-event probability therefore requires a small
$\Lambda$. The aggregate sampler instead permits $\Lambda\gtrsim1$ when
fractional cooling, kernel variation, orbit error and boundary risk are
controlled. This can save radiation updates in addition to photon draws.
The admissible trajectory step sets the final limit in a particle calculation.

The aggregation-capacity estimates above quantify the same opportunity:
for a fixed mean-cooling budget, the permitted photon load grows as
$\chi^{-1}$ in the soft ultrarelativistic limit. Kernel variation, orbit
resolution and boundary risk remain independent controls.

The lower-energy frozen-channel control below gives a dynamics speedup
of about 1.24 at approximately 1.91 explicit photons per nonempty step.
The separate corrected magnetic cooling test starts at
$\Lambda\simeq21.6$, demonstrating a larger aggregation opportunity without
assigning that test the channel timing ratio.

\section{Ion channel: directed radiation and separate timing controls}
The sole dimensionless field is $E_x=\widehat\omega_p^2\xi_x/2$, where $\widehat\omega_p=\omega_p\tc$ and $\xi_x=x/(c\tc)$. This corresponds to the physical channel field $\mathcal E_x=m\omega_p^2x/(2|e|)$. Both channel ensembles use
$n_e=3\times10^{17}\,\mathrm{cm}^{-3}$ in a prescribed broad linear focusing
field. The directed example compares corrected Gamma with state-updated
explicit sampling. The lower-energy example separately tests integrated
sources and the cost of frozen sampling.

\subsection{Symmetric-start directed-emission illustration}
The main-text directed-emission illustration uses
$n_e=3\times10^{17}\,\mathrm{cm}^{-3}$, $|x_0|=180\,\mu\mathrm m$,
$\gamma_0=30000$,
$\chi_0=0.01107626$, and 252 fixed steps over one initial betatron period.
The period is approximately 49.81 ps and the step is 0.19765 ps. The
initial local frozen photon load from the common CCF table is
$W(x_0)h\simeq0.591$; this is an initial value, not a measured mean over
the evolving ensemble. This illustration tests the directed source rather
than high-multiplicity aggregation.
There are 2000 electrons per model: 1000 start at each of the opposite turning
points $x_0=\pm180\,\mu\mathrm m$, with $p_{x0}=0$ and forward longitudinal
momentum. Each history has an independent random seed. No trajectory is
created by reflecting another, and no map is averaged with its mirror.
All histories complete and no source updates are flagged. The eight saved
paths per model comprise four starts on each side; color denotes the local
$\chi$, while the background increases with the harmonic potential
$U/(mc^2)=\widehat\omega_p^2\xi_x^2/4$.

The physical projected angle is
$\theta_x=\operatorname{atan2}(p_x,p_z)$. The number map represents
\begin{equation}
 \frac{1}{N_{e,0}}\frac{\dd^2N_\gamma}
 {\dd\ln E_\gamma\,\dd\theta_x},
\end{equation}
with $\theta_x$ in mrad. Its Gamma panel uses the corrected conditional mean
source, and its explicit panel counts accepted photons. Both use a common
absolute color scale. Smoothness results from conditional averaging over the
photon distribution, while electron trajectories remain stochastic; it is not
evidence of increased accuracy in a sparsely populated tail.

The sector geometry is a display transformation. Physical angles
$[-120,120]$ mrad map linearly to a drawn opening of $[-70,70]$ degrees.
Photon energy maps to radius as
\begin{equation}
 r=0.28+0.72\frac{\ln(E_\gamma/0.01\,\mathrm{MeV})}
 {\ln(1500/0.01)},
\end{equation}
with the visible range $0.01$--$1500$ MeV. Color retains density per physical
$\dd\ln E_\gamma\,\dd\theta_x$, not per unit sector area. There is no
solid-angle normalization or renormalization to the visible range. Integrals
use the original physical bin widths. No interpolation or smoothing is applied.
The full angular marginals reproduce the corresponding one-dimensional
spectra before display cropping.

As an integrated symmetry check, the relative negative/positive-angle energy
imbalance is $9.75\times10^{-4}$ for corrected Gamma and
$-2.43\times10^{-3}$ for state-updated explicit sampling, corresponding to
1.37 and $-0.51$ standard errors. Errors are estimated over independent
electrons within each fixed-size initial stratum, retaining correlations
between the two angular contributions from a given electron. This checks
integrated balance, rather than equality in every cell. The separate
$\gamma_0=20000$ timing and photon-load control below uses a different ensemble.

\subsection{Lower-energy aggregation and timing control}
The lower-energy controls use the prescribed channel field specified above. Parameters are $n_e=3\times10^{17}\,\mathrm{cm}^{-3}$, $\gamma_0=20000$, $x_0=180\,\mu$m, $p_{x0}=0$, and $p_{z0}=\sqrt{\gamma_0^2-1}$. The initial betatron parameter is $K=\gamma_0\omega_{\beta0}x_0/c=1855.3$. One initial period lasts 40.67 ps, or 12.19 mm of light propagation. There are 63 steps of .6455 ps and $10^4$ electrons per model. The prescribed channel isolates betatron motion and radiative damping in a
broad linear focusing field.
Explicit frozen sampling produces 786410 photons, or 1.2483 per external
step per initial electron ($10^4\times63$ steps); Gamma makes 411580 positive updates. The Poisson maximum is 12 photons per step. The largest realized step loss is 10.79\% (Gamma) and 10.70\% (Poisson), and the largest individual photon fraction is 8.37\%. The run is predominantly soft, with no imposed soft-photon cutoff. All paths complete. Empirical CDF distances are .0088 (energy), .0110 (transverse position), .0100 (transverse momentum), and .0116 (signed angle). Relative energy-budget residuals are below $9.3\times10^{-9}$. These statistics quantify agreement on the shared radiation grid.

The photon diagnostics use Sec.~\ref{sec:source}. Frozen Gamma uses the
first-order conditional source, corrected Gamma uses the second-order source,
and explicit spectra count accepted photons. All trajectories and all
source-admissibility checks complete in the corrected lower-energy control.

\begin{table}[tbp]\centering
\caption{Frozen-closure photon-count reference from full-kernel postprocessing, per initial electron; uncertainties are one standard error.}
\begin{tabular}{lrr}\toprule
Range & Gamma source & Explicit photons\\\midrule
All energies & $78.5042\pm.0036$ & $78.6410\pm.0881$\\
$E>100$ MeV & $4.13205\pm.00324$ & $4.12570\pm.01820$\\
$E>300$ MeV & $.207221\pm.000340$ & $.202200\pm.004317$\\
$E>1$ GeV & $2.64\times10^{-5}$ & 0 observed\\\bottomrule
\end{tabular}\end{table}

Full normalization includes the underflow below 1 keV. The state-updated
explicit control completes all $10^4$ histories and emits 785900 photons,
or 1.2475 per external step per initial electron. These counts concern the
full completed ensemble, not just nonempty steps.
The lower-energy benchmark reported here is integrated over direction.
The symmetric-start calculation above supplies the directed-emission
illustration with a finer fixed time grid.

With bulk output disabled, three alternating four-thread dynamics runs take .161388, .156785, and .181854 s for Gamma, versus .199658, .198569, and .210849 s for Poisson. The median ratio is 1.237. Compilation, tables, Python compression, plotting, and spectral postprocessing are excluded. Both models use a common coefficient routine that evaluates moments and
Gamma risk. The measured ratio characterizes this implementation and timed
region. It is not a timing of corrected Gamma against the state-updated
reference or of the directed-emission diagnostic.

\begin{figure}[tbp]
\centering\includegraphics[width=\textwidth]{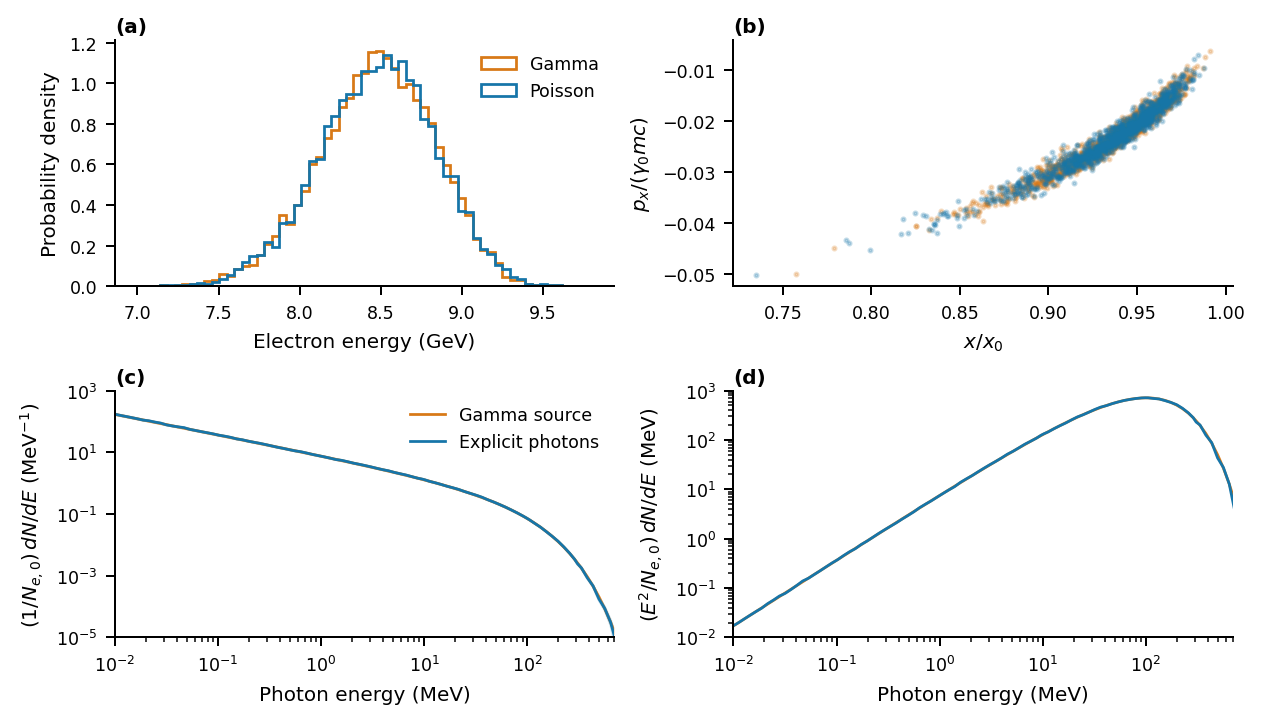}
\caption{Ion-channel electron distributions and photon sources. The full
kernel is evaluated along stochastic aggregate trajectories for the Gamma
source. Shading on the number spectrum is one estimated standard error over
electrons. Smoothness follows from conditional averaging.}
\end{figure}

\begin{table}[tbp]\centering\small
\caption{Frozen and corrected Gamma channel controls, with the frozen explicit
comparator: endpoints and binned mean-source energy, with
$10^4$ initial electrons per model. Endpoint errors are one standard error.
All trajectories and all source-admissibility checks complete. The binned
source energies are mean-source estimates per initial electron; realized
losses are sampled trajectory totals per initial electron.}
\begin{tabular}{lrrrr}\toprule
Sampler & Source order & $\langle E_e\rangle$ (GeV)
& Source energy (MeV) & Realized loss (MeV)\\\midrule
Gamma & 1 & $8.47550\pm0.00355$ & 1749.901 & 1750.138\\
Frozen explicit & 1 & $8.47486\pm0.00355$ & 1749.890 & 1750.780\\
Corrected Gamma & 2 & $8.48190\pm0.00352$ & 1743.593 & 1743.697\\
\bottomrule\end{tabular}\end{table}

\section{Focused counterpropagating beams}
\subsection{Field specification}
The beam construction follows Ref.~\cite{Elkina2014}, with common complex polarization $(1,i,0)$ and opposite focal axes. To fix focusing conventions, we specify the simulated focal-plane field. Put
\begin{equation}
 \ell=i\bar x-\bar y,\quad A=a_0\exp[-(\bar x^2+\bar y^2)/(2\bar b)]\ee^{-i\bar t},\quad C=iA.
\end{equation}
On $\bar z=0$,
\begin{align}
 \bar E_x&=\operatorname{Re}\{C(1+\ell\bar y/\bar b)\},\quad
 \bar E_y=\operatorname{Re}\{C(i-\ell\bar x/\bar b)\},\\
 \bar B_z&=\operatorname{Re}\{A\ell[(\bar x^2+\bar y^2)/\bar b^2-4/\bar b]\}.
\end{align}
Other components vanish on the simulated plane. Here
$\bar{\mathbf r}=\omega_0\mathbf r/c$ and $\bar t=\omega_0t$.
The barred fields are related to the dimensionless natural fields above by
$\mathbf E=\eta\bar{\mathbf E}$ and
$\mathbf B=\eta\bar{\mathbf B}$, where $\eta=\omega_0\tc$. We use combined $a_0=1000$ (500 per beam), $\bar b=50$, and $\hbar\omega_0=1$ eV, or $\lambda_0=1.23984\,\mu$m. There is no temporal envelope.

All electrons start at $\bar t=0$, $\bar x=-40$ and $\gamma_0=1000$,
with $p_x=\sqrt{\gamma_0^2-1}$ and $p_y=p_z=0$, and are followed to
$\bar t=110$. The main signed-impact ensemble tests the risk diagnostic
and radiation maps; the nine-impact ensemble supplies additional
matched-grid controls.

\subsection{Impact--angle emission maps and risk-colored trajectories}
The main-text focused illustration uses the same field and initial
energy, with 41 signed impacts $\bar y=-20,-19,\ldots,20$ and 256 independent
electrons per impact and model. There are 4400 fixed external steps to
$\bar t=110$, so $\Delta\bar t=0.025$ and each model has 10496 initial
histories. This ensemble is distinct from the nine-impact endpoint controls
below. The corrected Gamma sample has 93 excessive-loss stops and no
invalid-moment, midpoint, or model-exception stops; the state-updated explicit
sample completes. The accumulated active-Gamma risk is 93.53, compared with
93 observed Gamma stops.

The emission angle is $\theta=\operatorname{atan2}(p_y,p_x)$, relative to
the incoming $+x$ direction. The displayed number and energy yields at each
impact are normalized to the initial 256 electrons and to the angular bin
width in degrees. No impact-bin width, input beam profile, or solid angle is
included. The corrected Gamma panel accumulates the conditional sources in
Sec.~\ref{sec:source}; the explicit panel counts accepted photons.
The hard-photon diagnostic integrates bins above 100 MeV. Emission recorded
before a stop is retained, while the uncomputed continuation contributes
nothing; normalization is not changed to the survivor count. Thus these maps
compare recorded yields and do not impute the missing radiation of failed
histories.

Electron endpoints are recorded at the finite final time $\bar t=110$.
Some histories remain in the interaction region, so the endpoint maps are
not universally asymptotic scattering distributions.

The trajectory illustration replays fixed particle IDs at selected signed
impacts in both samplers. Gamma colors give the local conditional probability
$p_{\Gamma,\mathrm{over}}$ before the attempted draw. Two additional failed
histories are deliberately selected to illustrate boundary loss; their crosses
are located at the attempted radiation midpoint. The selected trajectories
are not a frequency sample. A central Lorentz-only trajectory provides the
no-radiation comparison. The field background is a geometric guide and is not
the risk color scale. The inset pools electron endpoints over all 41 equally
populated impacts, using the full initial count as denominator. The
impact-conditioned electron distributions below retain the dependence hidden
by that pooled inset; the main-text photon maps are not repeated here.

\begin{figure}[tbp]
\centering\includegraphics[width=\textwidth]{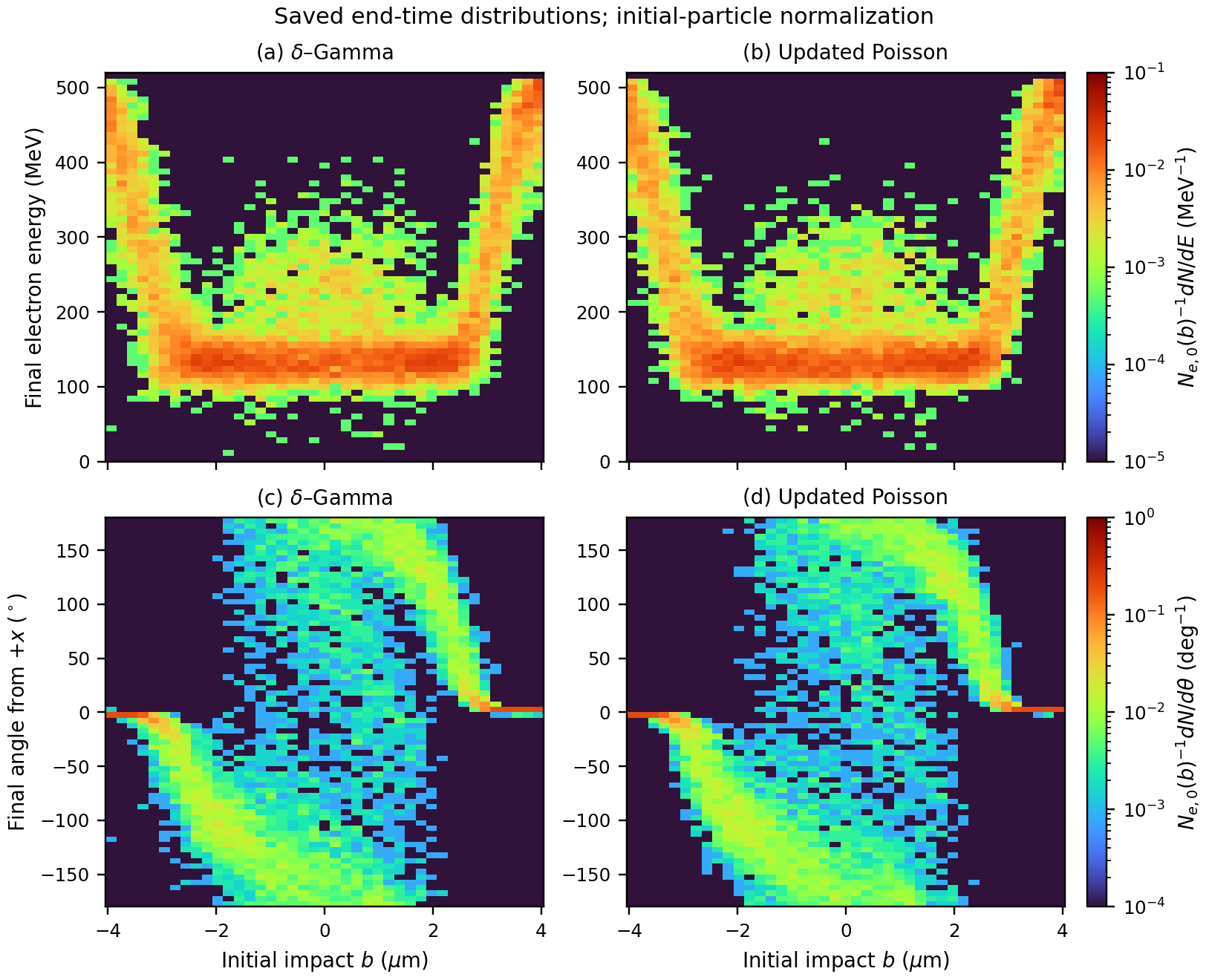}
\caption{Finite-end-time electron distributions versus signed initial impact
for the signed-impact ensemble. Columns compare corrected Gamma and
state-updated explicit sampling. Each impact is normalized to its initial
particle count, so failed histories leave missing endpoint mass rather than
renormalizing the surviving distribution. These are not beam-weighted or
universally asymptotic scattering spectra.}
\label{fig:focused_endpoints}
\end{figure}

\subsection{Additional matched-grid endpoint and switching controls}
Electrons start at $\bar t=0$ with
$p_x=\sqrt{\gamma_0^2-1}$ and $p_y=p_z=0$. Initial conditions are $\bar x=-40$, $\gamma=1000$, $p_y=0$, with impacts $\bar y=0,2,4,6,8,10,12,16,20$; final time is $\bar t=110$. Each impact has 2048 electrons per model. The endpoint control compares frozen and corrected closures, both hybrids,
and explicit references on the same two grids.

\begin{table}[tbp]\centering
\caption{Excessive-loss stops out of 18432 initial trajectories in the
matched-grid endpoint control. Each hybrid threshold is applied before a draw. All runs have
zero invalid-moment stops, midpoint failures and model exceptions; the
corrected hybrid makes zero invalid-moment fallbacks. Bernoulli controls
retain at most one emission per step.}
\begin{tabular}{lrr}\toprule
Sampler & $\Delta\bar t=.025$ & $\Delta\bar t=.0125$\\\midrule
Gamma & 217 & 235\\
Corrected Gamma & 202 & 230\\
Frozen Poisson & 9 & 6\\
State-updated explicit & 0 & 0\\
Hybrid, $10^{-5}$ & 59 & 93\\
Hybrid, $10^{-6}$ & 9 & 10\\
Corrected hybrid, $10^{-6}$ & 12 & 10\\
Bernoulli, $1-\ee^{-\Lambda}$ & 0 & 0\\
Bernoulli, $\Lambda$ & 0 & 0\\
\bottomrule\end{tabular}\end{table}

Excessive-loss stops do not decrease on these two grids, consistent with
the residual small-step risk in Eq.~\eqref{eq:small_step_risk}, although
the counts do not establish a monotonic trend. Cooling correction and
available-energy risk remain separate controls. The corrected hybrid reduces these failures to the same small-count
range as the frozen explicit and original hybrid controls.

The linear Bernoulli control checks $\Lambda\leq1$. These controls retain
zero or one event per step. Across both grids and impact groups, their mean
energy and angle remain within 1.56 and 1.87 combined standard errors of the
corresponding Poisson means. Boundary-event statistics are recorded separately
in the table.
Seven coarse Poisson failures contain two photons and two contain three. Their local $\chi$ range is .23--.52 and $\Lambda$ range .052--.103. Individual photons obey the cutoff; their frozen sum does not. The frozen-packet energy budget is the mechanism tested by this control.
For the $10^{-5}$ hybrid, accumulated active-Gamma risk grows from 44.86 to 88.02 under refinement; for $10^{-6}$ it grows from 2.50 to 6.23. More steps and a changed conditional shape modify total risk even with an unchanged threshold. Histograms use initial-ensemble normalization; endpoint means refer to survivors.

\section{Positive support and the frozen diffusion comparison}
The illustrative density in the main text uses the classical soft-synchrotron
photon-number law, with $y=\epsilon/\epsilon_c$ and
$\epsilon_c=3\gamma\chi/2$:
\begin{equation}
 f(y)=\frac3{5\pi}\int_y^\infty K_{5/3}(u)\,\dd u.
\end{equation}
Write $C_n=\int_0^\infty y^nf(y)\,\dd y$; these normalized moments are
derived in Supplement II. For $Z=S/\epsilon_c$,
the compound-Poisson mean and variance are
$\Lambda C_1$ and $\Lambda C_2$, respectively. The frozen drift--diffusion
comparison uses a Gaussian with exactly those two moments. Its probability of
a radiative gain is
\begin{equation}
 \Pr_{\rm Gauss}(Z<0)
 =\Phi\!\left(-\frac{\sqrt\Lambda\,C_1}{\sqrt{C_2}}\right),
\end{equation}
where $\Phi$ is the standard-normal CDF. At $\Lambda=5$, this probability is
0.14036. The positive compound-Poisson and delta--Gamma laws assign zero
probability to that region. Both retain the no-emission mass
$\exp(-5)=0.006738$. The comparison identifies a structural distinction between
finite-step laws with identical moments.

The compound-Poisson curve is evaluated by a zero-padded FFT convolution on a
left-edge lattice. The final grid has $2^{20}$ cells over $0\leq Z<48$;
its doubled FFT domain suppresses wraparound. Halving the grid spacing changes
the CDF by at most $3.09\times10^{-5}$ over $0.01\leq Z\leq25$.
The first four FFT moments agree with the compound-Poisson moments of the
lattice photon law to a maximum relative error of $7.3\times10^{-12}$.
The lattice discretization gives a coupled first-moment error bounded by
$\Lambda\Delta Z$. The density panel displays the continuous part; the CDF
includes the zero atom explicitly.

\subsection{Auxiliary shifted-Gamma comparison}
A shifted Gamma component fitted to three conditional moments provides an
additional support control, rather than a variant of the proposed update. The fitted moments in this comparison are those of $Z$. With positive-branch third central moment $\mu_{3,+}$,
its parameters are
\begin{equation}
 \theta_3=\frac{\mu_{3,+}}{2v_+},\qquad
 k_3=\frac{v_+}{\theta_3^2},\qquad
 a_3=\mu_+-k_3\theta_3.
\end{equation}
For the same example, $a_3=-0.18090$ in scaled energy and the mixture assigns
probability 0.04420 to $Z<0$. Thus adding a moment through an unconstrained
shift can sacrifice physical support. The delta--Gamma construction imposes
that support directly.
\begin{figure}[tbp]
\centering\includegraphics[width=\textwidth]{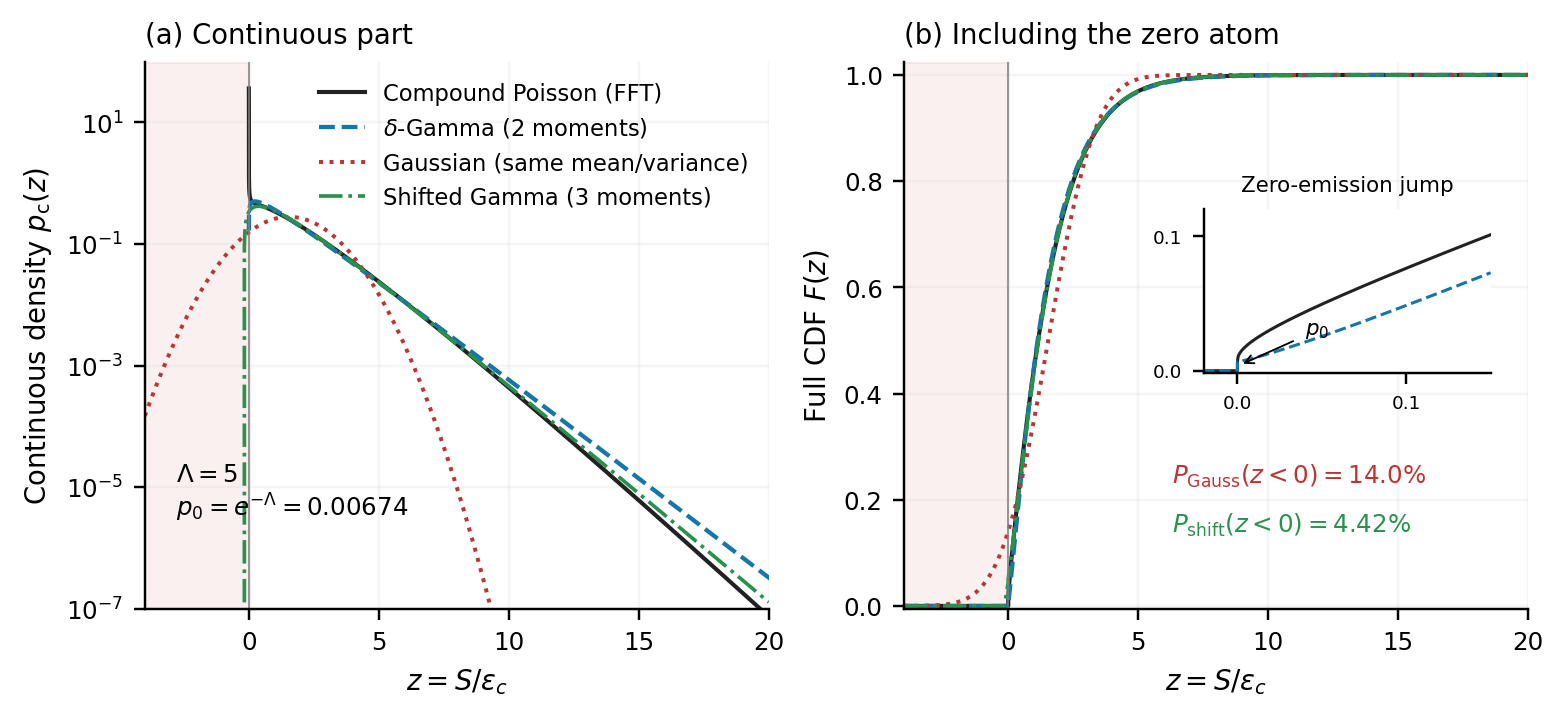}
\caption{Frozen-step density and CDF at $\Lambda=5$, including a three-moment
shifted Gamma comparison. Shading marks negative loss. The zero atom is shown
as a CDF jump, with an inset magnifying the region near zero. The Gaussian
curve is the frozen two-moment diffusion increment.}
\end{figure}

\section{Construction and practical scope}
The construction replaces individual soft-photon draws by a positive
aggregate update on a resolved particle trajectory. State feedback corrects
its selected finite-step moments; Supplement~II establishes the resulting
cooling improvement independently of agreement between frozen samplers.

The directed channel example connects this update to a mean radiation
source. Focused scattering tests the excessive-loss diagnostic, with stops
and recorded yields accounted for at the initial-ensemble normalization.
The separate frozen-channel control quantifies implementation cost on a
common grid. Together these tests establish distinct functions of the
aggregate method within the adopted local kinetic model.

\bibliographystyle{apsrev4-2}\bibliography{references}


\title{Supplement II: Synchrotron analysis and magnetic verification}

\author{Nina Elkina$^1$ and Alexander Pukhov$^{1}$}
\affiliation{$^1$Institut f\"ur Theoretische Physik I, Heinrich-Heine-Universit\"at D\"usseldorf, 40225 D\"usseldorf, Germany}

\date{November 19, 2026}
\maketitle
\section{Purpose, setups, and units}
Independent analytical cooling moments and a state-updated photon reference
together identify frozen-state bias and test its reduction by corrected
aggregate moments. Moment normalization and kernel checks establish the
inputs, while separate orbit and diffusion comparisons test trajectory
resolution and physical support. All comparisons use the spin-averaged kernel and
collinear recoil convention of Supplement I.

We set $\mathbf E=0$, $\mathbf B=b\mathbf e_z$ and initially transverse
momentum, with $b=B/B_{\rm cr}$. The three setups have distinct roles:
\begin{itemize}
\item Energy-only cooling: $\gamma_0=2000$ and
$\chi_0=0.002,0.01,0.02$, evolved to scaled cooling time $u=1$.
This isolates recoil feedback from orbit integration.
\item Orbit aggregation: $b=10^{-6}$ and $\gamma_0=1000$, evolved
for five initial gyroperiods to test the trajectory-imposed step limit.
\item Diffusion support: $\gamma_0=1000$ and $\chi_0=0.04$, evolved
for one initial gyroperiod to visualize the initial-energy boundary.
\end{itemize}
The energy-only state-updated control uses the first setup at
$\chi_0\simeq0.002$. Its explicit reference updates the emitter after each
accepted photon; the finer frozen-packet references are identified separately.

Rates are per natural laboratory time $\tau=t/\tc$, energies are in $mc^2$,
and momenta are in $mc$. The local quantum parameter is
$\chi=b\sqrt{\gamma^2-1}$. In the soft ultrarelativistic limit,
\begin{equation}
 W=\frac{5\alpha\chi}{2\sqrt3\gamma}[1+O(\chi)]
 \simeq\frac{5\alpha b}{2\sqrt3}.
\end{equation}
The leading magnetic rate is therefore approximately independent of energy.
All sampling calculations retain the full quantum rate and moments.

\section{Soft-recoil moments from the Airy representation}
\subsection{Equivalent kernel and conventions}
To derive the leading small-$\chi$ coefficients, keep the mass explicit while using $\hbar=c=1$. The rate $\mathcal W$ in the next equation is per dimensional natural laboratory time $t$; the rate $W$ elsewhere is per $\tau=mt$. Write the dimensional electron energy as $\mathcal E=\gamma m$ and the photon energy as $\omega=m\epsilon=\mathcal E s$. Define $\kappa=\chi s$, $\chi'=\chi-\kappa$, and
\begin{equation}
 x=\left(\frac{\kappa}{\chi\chi'}\right)^{2/3},\qquad
 s=\frac{\chi x^{3/2}}{1+\chi x^{3/2}},\qquad z=\frac23x^{3/2}.
\end{equation}
Here $x$ is the Airy argument; the distinct symbol $y$ below denotes photon energy scaled to the classical critical energy. The Airy form of the same uncut ultrarelativistic kernel is
\begin{equation}
 \frac{\dd\mathcal W}{\dd\omega}=-\frac{\alpha m^2}{\mathcal E^2}
 \left[\int_x^\infty\!\operatorname{Ai}(u)\dd u
 +\left(\frac2x+\kappa\sqrt x\right)\operatorname{Ai}'(x)\right].
 \label{eq:airy_kernel}
\end{equation}
The Bessel--Airy identities, together with $z=2x^{3/2}/3$, transform Eq.~\eqref{eq:airy_kernel} into the dimensional-time version of the kernel in Supplement I after $\dd\omega=\mathcal E\,\dd s$; dividing $\dd\mathcal W/\dd s$ by $m$ gives the rate per $\tau$. The same finite-energy cutoff can be imposed in either representation. For the asymptotic calculation we integrate $0<\omega<\mathcal E$ and take $\chi\to0$ at fixed ultrarelativistic $\gamma$; the difference due to the simulation cutoff is beyond the leading soft coefficients.

Define the dimensional moment rates
\begin{equation}
 M_n=\int_0^{\mathcal E}\omega^n\frac{\dd\mathcal W}{\dd\omega}\dd\omega,
 \qquad R_n=\frac{M_n}{m^{n+1}}\quad(\hbar=c=1).
\end{equation}
For physical moment rates $\widehat M_n$ per second, the conversion is $R_n=\tc\widehat M_n/(mc^2)^n$. In particular, $R_0=M_0/m=W$. For fixed $x=O(1)$,
\begin{equation}
 \omega=\mathcal E\chi x^{3/2}[1+O(\chi)],\quad
 \dd\omega=\tfrac32\mathcal E\chi x^{1/2}[1+O(\chi)]\dd x,
 \quad \kappa\sqrt x=O(\chi^2).
\end{equation}
Consequently, for integer $n\geq1$,
\begin{align}
 M_n&=\alpha m^2\mathcal E^{n-1}
 \left[-\tfrac32J_n\chi^{n+1}+O(\chi^{n+2})\right],\\
 J_n&=\int_0^\infty x^{(3n+1)/2}
 \left[\int_x^\infty\operatorname{Ai}(u)\dd u+
 \frac2x\operatorname{Ai}'(x)\right]\dd x.
 \label{eq:Jn}
\end{align}

\subsection{Mellin integral and the first four moment rates}
The required Airy Mellin integral is
\begin{equation}
 A_\nu=\int_0^\infty x^\nu\operatorname{Ai}(x)\dd x
 =\frac{3^{2\nu/3}}{2\pi\sqrt3}
 \Gamma\!\left(\frac{\nu+1}{3}\right)
 \Gamma\!\left(\frac{\nu+2}{3}\right),\qquad \nu>-1.
\end{equation}
Putting $a=(3n+1)/2$, interchange of integration and integration by parts give
\begin{equation}
 J_n=\frac{A_{a+1}}{a+1}-2(a-1)A_{a-2},\qquad n\geq1.
\end{equation}
The boundary term $x^{a-1}\operatorname{Ai}(x)$ vanishes for $n\geq1$. For $n=0$, the total rate is evaluated separately. The first values are
\begin{equation}
 J_1=-\frac49,\quad J_2=-\frac{55\sqrt3}{108},\quad
 J_3=-\frac{28}{9},\quad J_4=-\frac{1309\sqrt3}{144}.
\end{equation}
The moment rates per natural laboratory time, for $\epsilon=\omega/(mc^2)$, are
\begin{align}
 R_0&=\frac{5\alpha\chi}{2\sqrt3\gamma}[1+O(\chi)], &
 R_1&=\frac{2\alpha\chi^2}{3}[1+O(\chi)],\\
 R_2&=\frac{55\alpha\gamma\chi^3}{24\sqrt3}[1+O(\chi)], &
 R_3&=\frac{14\alpha\gamma^2\chi^4}{3}[1+O(\chi)],\\
 R_4&=\frac{1309\alpha\gamma^3\chi^5}{32\sqrt3}[1+O(\chi)].
 \label{eq:soft_rates}
\end{align}
Rates per laboratory second are obtained by dividing these expressions by $\tc$. The numerical experiments use the full quantum moments from the kernel in Supplement I.

\subsection{Single-photon normalization and an independent Bessel check}
Let $\epsilon_c=3\gamma\chi/2$, $y=\epsilon/\epsilon_c$, and $\omega_c=mc^2\epsilon_c$. The normalized classical photon-number density and its raw moments are
\begin{align}
 f_{\rm cl}(y)&=\frac3{5\pi}\int_y^\infty K_{5/3}(u)\dd u,\qquad
 \int_0^\infty f_{\rm cl}(y)\dd y=1,\\
 C_n&=\int_0^\infty y^nf_{\rm cl}(y)\dd y
 =\frac{3\,2^n}{5\pi(n+1)}
 \Gamma\!\left(\frac n2+\frac{11}{6}\right)
 \Gamma\!\left(\frac n2+\frac16\right).
\end{align}
These classical coefficients independently reproduce the Airy result through
$\mu_n=C_n(3\chi/2)^n[1+O(\chi)]$ and
$m_n=C_n\epsilon_c^n[1+O(\chi)]$.

\begin{table}[tbp]\centering
\caption{Leading single-photon moments in two energy normalizations. Entries in the middle column multiply $\chi^n$; entries in the last column multiply $\epsilon_c^n$ for $m_n$, or $\omega_c^n$ for $\langle\omega^n\rangle$. Relative quantum corrections are $O(\chi)$.}
\begin{tabular}{ccc}\toprule
$n$ & $\mu_n/\chi^n$ (leading) & $C_n$\\\midrule
1 & $4\sqrt3/15$ & $8/(15\sqrt3)$\\
2 & $11/12$ & $11/27$\\
3 & $28\sqrt3/15$ & $224/(135\sqrt3)$\\
4 & $1309/80$ & $1309/405$\\\bottomrule
\end{tabular}\end{table}

In particular, $11/12$ is the coefficient of $(\gamma\chi)^2$ in $m_2$; the coefficient of $\epsilon_c^2$ is $11/27$. Mixing these scales changes the loss variance.

\section{Independent kernel and frozen-packet checks}
The common table was compared with adaptive direct integration of the CCF
Bessel kernel, without using its generator or stored quadrature. At nine
sampled $\chi$ values from $10^{-5}$ to 1, the largest absolute relative
errors were approximately $9.2\times10^{-5}$ for $W$, $6.2\times10^{-5}$ for
$m_1$, and $1.5\times10^{-4}$ for $m_2$. These errors quantify the tested nodes. The endpoint was regularized with a cubic transformed
variable; tighter tolerances and a longer integration interval changed the
checked integrals by about $10^{-12}$ or less at $\chi=0.002$ and 1.

The finite-step compound-Poisson identity provides a separate distributional
reference in a frozen state:
\begin{equation}
 \mathbb E[\ee^{-sS}]
 =\exp\{\Lambda[\mathbb E(\ee^{-s\epsilon})-1]\},\qquad s\geq0.
\end{equation}
The single-photon expectation is computed by direct quadrature of the kernel.
At $\gamma=2000$, $\chi=0.002$ and $\Lambda=3$, a $3\times10^5$-packet
explicit sample agrees with this reference at all six checked arguments,
with discrepancies below 0.47 estimated standard errors.
The Gamma mixture preserves its prescribed zero mass and first two moments,
but its transform differs systematically: at $s\epsilon_c=1$ the exact frozen
transform is approximately 0.56429 and the moment-matched Gamma transform is
0.55901. At larger positive $s$ the transform emphasizes small losses. This
check resolves the shape difference of the frozen laws, with emphasis on small
losses at larger transform arguments. Evolving-state distributions are tested
separately below.

\section{Independent analytical magnetic-cooling reference}
The analytical reference distinguishes the physical quantum correction
from finite-step cooling error. Agreement of frozen Gamma with frozen
photon sampling alone cannot make this distinction. The energy moments
derived by Baier, Katkov, and Strakhovenko
\cite{BaierKatkovStrakhovenko1974} provide an independent analytical reference. Their Eq.~(2.31) gives the moments for
an initially monoenergetic ultrarelativistic beam transverse to a uniform
magnetic field, as an expansion in the initial quantum parameter.
In the notation of that reference, the expansion parameter $\chi_0$ and the
scaled time $z$ correspond to
\begin{equation}
 \chi_0=b\gamma_0,\qquad
 z=u=\frac{\mathcal I_c(E_0)t}{E_0}
 =\frac23\alpha b^2\gamma_0\tau,
 \quad \tau=\frac{t}{\tau_C},\quad E_0=mc^2\gamma_0.
\end{equation}
Here $\mathcal I_c(E_0)$ is the classical radiation power at the initial
energy, $b=B/B_{\rm cr}$, and $\tau_C=\hbar/(mc^2)$.
This is laboratory time. The equality uses the ultrarelativistic transverse
classical power; our sampling code instead evaluates
$\chi=b\sqrt{\gamma^2-1}$, whose difference from $b\gamma$ is negligible in
these tests. We rename the source's moment exponent $m$ as $\nu$ to distinguish
it from the electron mass. The source's intermediate expansion (2.23)
uses $(3\chi_0)^k$; its final Eq.~(2.31), used here, is already expressed
in powers of $\chi_0$.

Let $H=1+u$, $L=\ln H$, and $d=55\sqrt3/48=55/(16\sqrt3)$.
Equation~(2.31) of Ref.~\cite{BaierKatkovStrakhovenko1974}, translated into our
normalization and retained through second order, is
\begin{align}
 \frac{\langle\gamma^\nu\rangle}{\gamma_0^\nu}
 =H^{-\nu}\Bigg\{1
 &+\chi_0\frac{d\nu}{H}
 \left[2L+\frac{(\nu+1)u}{2H}\right]\nonumber\\
 &+\chi_0^2\frac{\nu}{H^2}
 \left[-P_\nu(u)+d^2Q_\nu(u)\right]\Bigg\}
 +O(\chi_0^3),\label{eq:baier_general_moments}
\end{align}
where
\begin{align}
 P_\nu(u)&=27u+\frac{7u(\nu+1)(\nu+2)}{6H}
             +7(\nu+1)L,\\
 Q_\nu(u)&=2(\nu+1)L^2
 +\frac{L}{H}\left[\nu(\nu+5)u+2(\nu-1)\right]\nonumber\\
 &\quad+\frac{u}{H^2}\left[5u^2
 +\frac{u}{8}(62-5\nu+6\nu^2+\nu^3)-2(\nu-1)\right].
\end{align}
The expansion is taken at fixed $u$ and retained through $\chi_0^2$.
The source's additional third-order mean correction is given in its Eq.~(2.32).

For the mean and central variance, it is useful to write
\begin{align}
 \frac{\langle\gamma\rangle}{\gamma_0}
 &=a_0+\chi_0a_1+\chi_0^2a_2+O(\chi_0^3),\\
 \frac{\operatorname{Var}\gamma}{\gamma_0^2}
 &=\chi_0v_1+\chi_0^2v_2+O(\chi_0^3),
\end{align}
with
\begin{align}
 a_0&=H^{-1},\qquad
 a_1=\frac{d}{H^2}\left(2L+\frac{u}{H}\right),\\
 a_2&=\frac1{H^3}\Bigg[-27u-\frac{7u}{H}-14L\nonumber\\
 &\hspace{17mm}+d^2\left(4L^2+\frac{6uL}{H}
                  +\frac{u^2(5u+8)}{H^2}\right)\Bigg],\\
 v_1&=\frac{du}{H^4},\\
 v_2&=\frac1{H^4}\Bigg[-14\left(L+\frac{u}{H}\right)\nonumber\\
 &\hspace{17mm}+4d^2\left(\frac{(3u+1)L}{H}
                   +\frac{u(u-1)}{H^2}\right)\Bigg].
\end{align}
The variance is consistently expanded to the same order:
if $F_{\nu,j}$ denotes the coefficient of $\chi_0^j$ in
Eq.~\eqref{eq:baier_general_moments}, then
$v_2=F_{2,2}-2F_{1,0}F_{1,2}-F_{1,1}^2$.
Squaring a truncated mean and retaining its induced third- and fourth-order
terms would introduce uncontrolled contributions.

The first quantum correction to the ensemble mean differs from integrating
the quantum-corrected power along a deterministic trajectory. Such an
integration gives
\begin{equation}
 \frac{\gamma_{\rm det}}{\gamma_0}
 =H^{-1}+\chi_0\frac{3dL}{H^2}+O(\chi_0^2).
\end{equation}
Its difference from the ensemble mean is accounted for by stochastic energy
spread. Thus quantum power suppression and the feedback of fluctuations on
cooling must both enter the analytical comparison.

At $u=1$ and $\gamma_0=2000$, the successive analytical values are
\begin{center}
\begin{tabular}{c|rrr}
 $\chi_0$ & $\langle\gamma\rangle_{O(\chi_0)}$
 & $\langle\gamma\rangle_{O(\chi_0^2)}$
 & $\operatorname{Var}_{O(\chi_0^2)}/\operatorname{Var}_{\rm leading}$\\\hline
 0.002 & 1003.744 & 1003.732 & 1.00518\\
 0.01  & 1018.718 & 1018.427 & 1.02589\\
 0.02  & 1037.436 & 1036.272 & 1.05177
\end{tabular}
\end{center}
The approximately five-percent variance departure from the leading formula
at $\chi_0=0.02$ is therefore predicted by the next analytical order. The second-order reference separates this physical quantum contribution from
numerical cooling bias.

\subsection{Independent algebraic checks}
We additionally verify the translated coefficients against the small-$\chi_0$
expansion of the backward jump generator. For $x=\gamma/\gamma_0$ and time $u$,
\begin{align}
 \mathcal G_0f&=-x^2f',\\
 \mathcal G_1f&=3dx^3f'+\tfrac12dx^4f'',\\
 \mathcal G_2f&=-48x^4f'-14x^5f''-\tfrac76x^6f''',
\end{align}
where $\mathcal G=\mathcal G_0+\chi_0\mathcal G_1
+\chi_0^2\mathcal G_2+\cdots$.
For moment powers one and two, the implemented coefficients satisfy the
backward evolution equations and the initial conditions through second order
symbolically. The third derivative is part of this perturbative coefficient check; the
simulations sample the full kernel rather than a finite Kramers--Moyal
truncation.

\begin{figure*}[t]
\centering
\includegraphics[width=\textwidth]{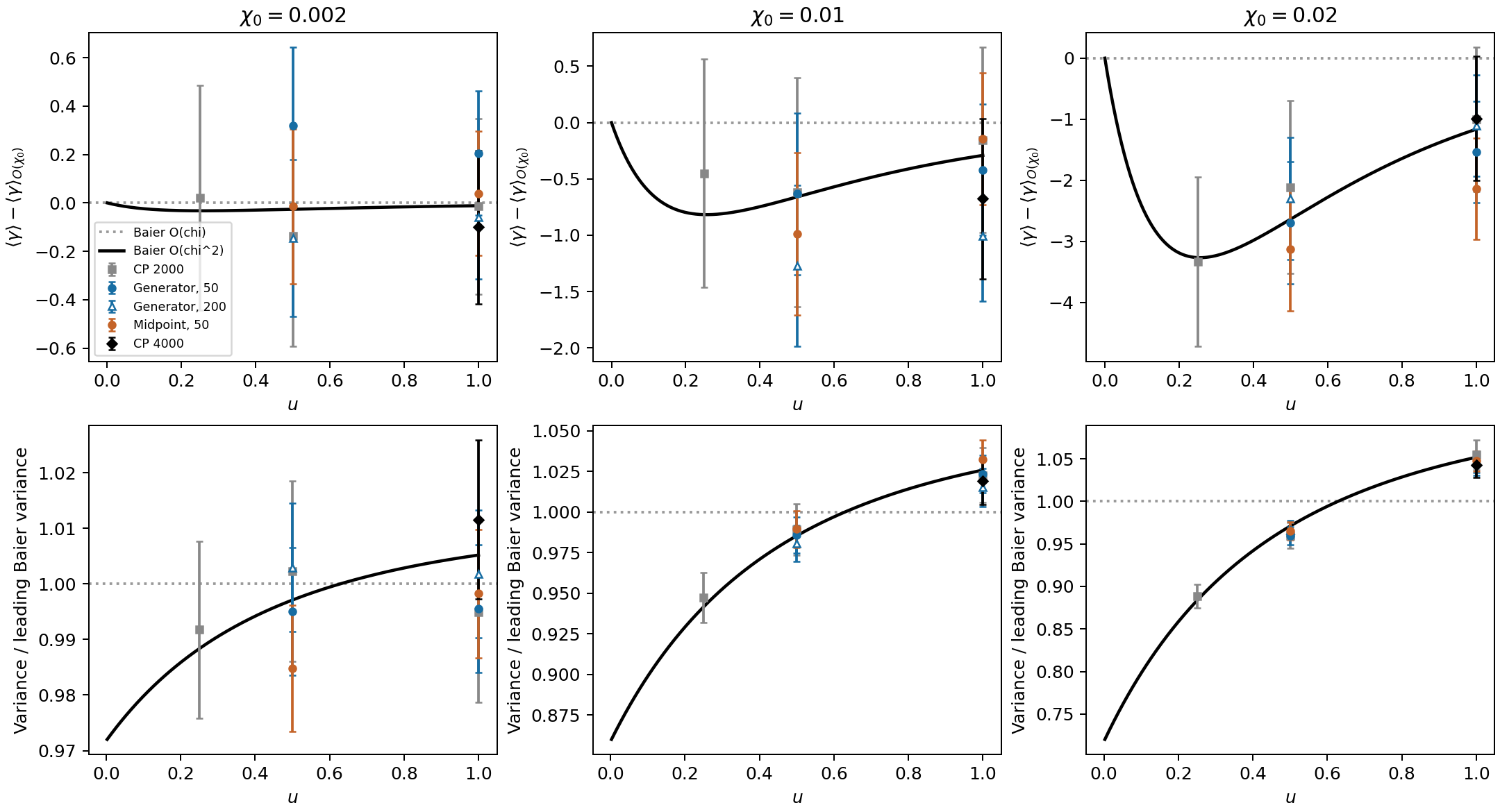}
\caption{Magnetic cooling compared with the analytical moments of
Baier, Katkov, and Strakhovenko, through first and second order in $\chi_0$.
Top: mean energy relative to the first-order prediction.
Bottom: variance normalized by its leading analytical term.
The dotted baselines are the first-order predictions; black curves include
the second-order terms. Independent explicit compound-Poisson ensembles use
2000 or 4000 steps. Corrected aggregate ensembles use 50 or 200 steps.
Error bars show twice the estimated Monte Carlo standard error.
The two analytical orders display the reference's truncation dependence.}
\label{fig:baier_moment_reference}
\end{figure*}

\section{Finite-step cooling-feedback correction}
Let the starting energy be $g$. At fixed local field, let
$w(g,\epsilon)$ denote the differential emission rate per photon energy and
per unit natural laboratory time, with $0<\epsilon<g-1$. Because magnetic rotation leaves the energy and $\chi$ unchanged between
emissions, the radiation calculation closes on energy. The backward
one-photon jump generator and its first two positive moment rates are
\begin{align}
 (\mathcal Lf)(g)&=\int_0^{g-1}
 [f(g-\epsilon)-f(g)]w(g,\epsilon)\,\dd\epsilon,\\
 A(g)&=\int_0^{g-1}\epsilon w(g,\epsilon)\,\dd\epsilon=R_1(g),\\
 B(g)&=\int_0^{g-1}\epsilon^2w(g,\epsilon)\,\dd\epsilon=R_2(g).
\end{align}
Thus $B$ is a raw second recoil-moment rate; the diffusion coefficient in
an operator written as $\partial_g^2(Df)$ would be $D=B/2$.
The semigroup expansion is
\begin{equation}
 \mathbb E_g[f(\gamma_h)]
 =f(g)+h\mathcal Lf(g)+\tfrac12h^2\mathcal L^2f(g)+O(h^3).
\end{equation}
Taking $f(y)=y$ gives $\mathcal Lf=-A$ and hence
\begin{equation}
 M=\mathbb E_g[g-\gamma_h]=hA+\tfrac12h^2\mathcal LA+O(h^3).
\end{equation}
For the second loss moment use $f_g(y)=(g-y)^2$, with $g$ held fixed.
Then $\mathcal Lf_g(y)=2(g-y)A(y)+B(y)$, so
\begin{equation}
 \mathbb E_g S^2=hB+\tfrac12h^2
 \left[\mathcal LB+2\int\epsilon A(g-\epsilon)w(g,\epsilon)\dd\epsilon\right]
 +O(h^3).
\end{equation}
Subtracting $M^2$ yields
\begin{align}
 V&=hB+\tfrac12h^2\mathcal LB+h^2C+O(h^3),\\
 C&=\int\epsilon[A(g-\epsilon)-A(g)]w(g,\epsilon)\dd\epsilon.
\end{align}
The $h^2A^2$ contribution cancels when the variance is formed from the raw
second moment. If $A$ and $B$ were state independent, all three correction
coefficients would vanish and the frozen variance $hB$ would be recovered.

For positive $M,V$ and conditional variance, Supplement I's matching formula
constructs the corrected Gamma branch. In this uniform magnetic field a
nonemitting electron retains its energy and rate, so the full no-event
probability over natural duration $\tau_{\rm end}$ is
$\exp[-W(\gamma_0)\tau_{\rm end}]$. The local weights multiply to this
value along the nonemitting path, since $\sum_jh_j=\tau_{\rm end}$.
For varying fields or nonradiating energy, the rate must instead be
integrated along that path, as discussed in Supplement I.

\subsection{Cooling results and tested window}
The energy-only tests use $\gamma_0=2000$, $\chi_0=0.002,0.01,0.02$,
$b=\chi_0/\gamma_0$, and final scaled cooling time $u=1$.
Each aggregate run contains $6\times10^4$ independent electrons and 50, 100,
or 200 fixed steps. The baseline, generator correction and auxiliary midpoint/transport
comparator use independent,
reproducible random streams. The energy-only setup isolates cooling feedback
from orbit integration. At $\chi_0=0.002$, the 50-step run has an initial
local frozen load $\Lambda_0\simeq21.6$; 100 and 200 steps give
approximately 10.8 and 5.4, respectively. These are initial values of
$W(\gamma_0)h$, not measured photon counts averaged over evolving histories.

\begin{table}[tbp]\centering\small
\caption{Magnetic energy-only cooling at $u=1$. Quoted errors are one
Monte Carlo standard error of the mean. The second-order analytical means are
1003.732, 1018.427 and 1036.272 for the three fields.}
\begin{tabular}{crrrr}\toprule
$\chi_0$ & Steps & Frozen Gamma & Generator correction & Midpoint/transport\\\midrule
0.002 & 50 & $996.76\pm0.13$ & $1003.95\pm0.13$ & $1003.78\pm0.13$\\
0.002 & 100 & $1000.26\pm0.13$ & $1003.97\pm0.13$ & $1003.73\pm0.13$\\
0.002 & 200 & $1001.91\pm0.13$ & $1003.69\pm0.13$ & $1003.63\pm0.13$\\
0.01 & 50 & $1011.94\pm0.29$ & $1018.30\pm0.29$ & $1018.58\pm0.29$\\
0.01 & 100 & $1014.99\pm0.29$ & $1018.56\pm0.29$ & $1018.23\pm0.29$\\
0.01 & 200 & $1016.63\pm0.29$ & $1017.71\pm0.29$ & $1018.72\pm0.29$\\
0.02 & 50 & $1029.58\pm0.42$ & $1035.90\pm0.42$ & $1035.30\pm0.42$\\
0.02 & 100 & $1033.15\pm0.42$ & $1035.35\pm0.42$ & $1036.08\pm0.42$\\
0.02 & 200 & $1034.92\pm0.42$ & $1036.33\pm0.42$ & $1036.61\pm0.41$\\
\bottomrule\end{tabular}\end{table}

\begin{figure}[tbp]
\centering\includegraphics[width=\textwidth]{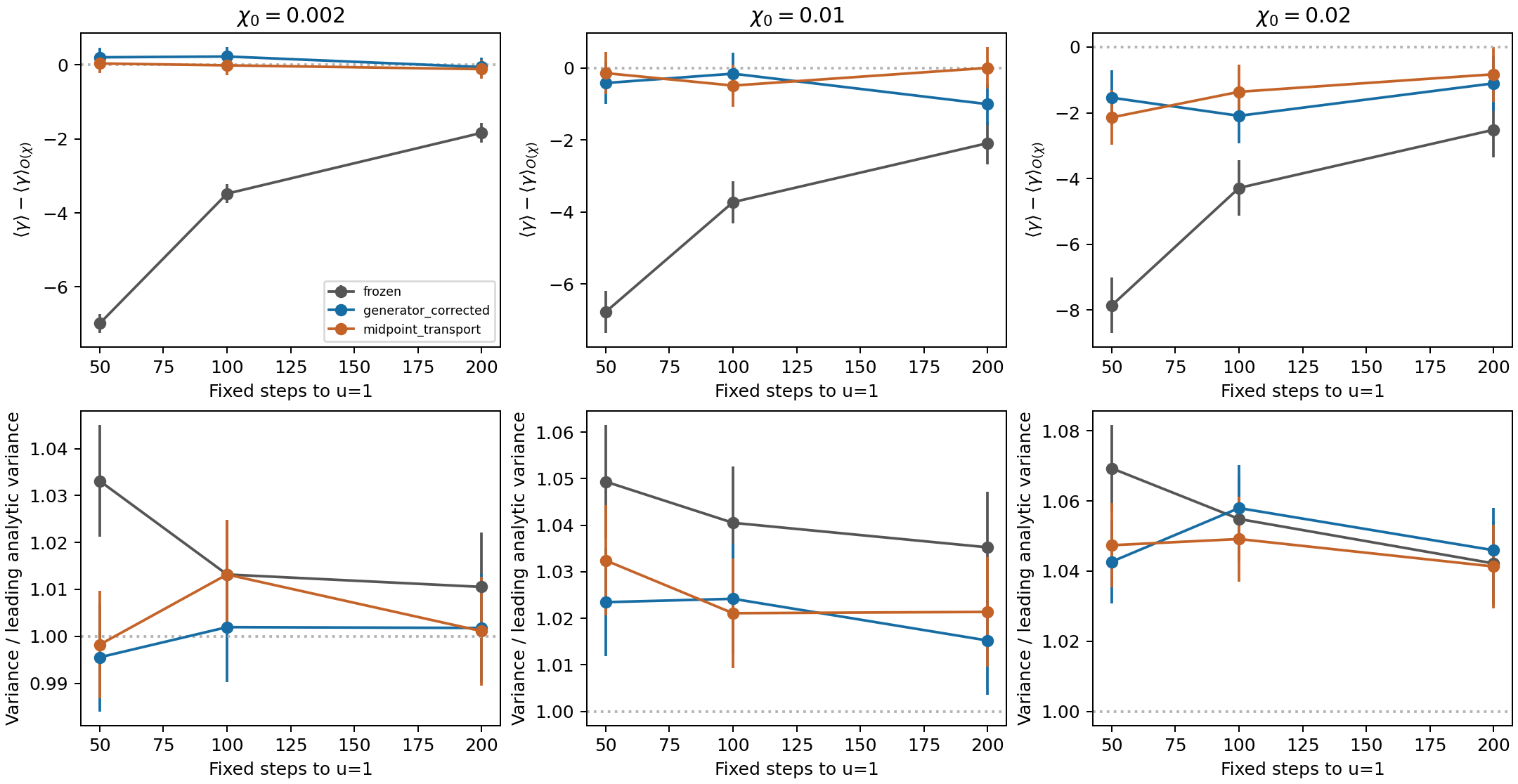}
\caption{Step dependence of the energy-only cooling test. Top: mean relative
to the first-order analytical reference. Bottom: variance divided by its leading
analytical term. Error bars are twice estimated Monte Carlo standard errors.
The residual second-order analytical contributions are shown explicitly in the
preceding analytical-reference figure.}
\end{figure}

\section{State-updated magnetic reference}
Uniformization provides a state-updated reference that changes the emitter after
each photon, while keeping the external radiation intervals fixed. We use
$\gamma_0=2000$, $b=10^{-6}$, $\chi_0\simeq0.002$ and cooling duration $u=1$.
Each row below uses $10^5$ independent trajectories. The quoted uncertainty is
the standard error of the mean. The second-order analytical mean is 1003.732;
its truncation and numerical kernel errors are distinct from sampling error.
\begin{table}[tbp]\centering
\caption{Magnetic energy-only control. $r$ multiplies the uniformization majorant.}
\begin{tabular}{lrrr}\toprule
Sampler & Steps & $r$ & Mean $\gamma$\\\midrule
Frozen explicit & 50 & -- & $996.694\pm0.101$\\
State-updated explicit & 50 & 1 & $1003.492\pm0.100$\\
State-updated explicit & 100 & 1 & $1003.487\pm0.100$\\
State-updated explicit & 200 & 1 & $1003.667\pm0.100$\\
State-updated explicit & 50 & 2 & $1003.614\pm0.100$\\
\bottomrule\end{tabular}\end{table}
The fixed-grid and doubled-majorant controls agree at the resolution of the
independent ensembles. In this energy-only magnetic test the underlying local
semigroup is independent of the partition; these checks test its implementation.
Against the 200-step state-updated ensemble, the empirical CDF distance of
50-step frozen Gamma is 0.08755, compared with 0.00568 for generator-corrected
Gamma and 0.00555 for the intermediate-state approximation. The Gamma
ensembles contain $6\times10^4$ trajectories each. These are finite-ensemble
distances, not uniform accuracy bounds. 
While the frozen explicit control identifies a bias shared with the
Gamma closure, the CDF comparison shows that feedback improves both the
mean and the energy distribution without resolving individual emissions.
Together these checks validate the correction against an unfrozen reference.
The fine-grid frozen references in the next section extend the distribution
comparison to three initial quantum parameters.

\begin{figure}[tbp]\centering
\includegraphics[width=\textwidth]{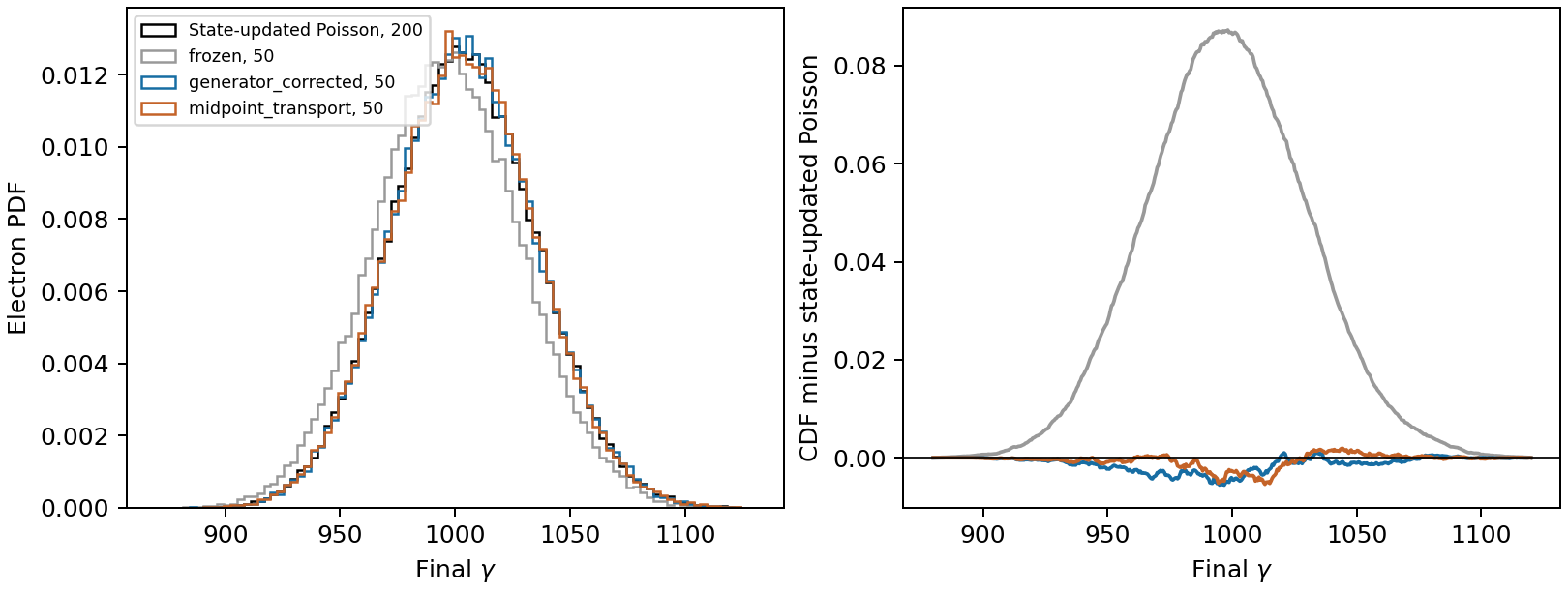}
\caption{Endpoint distributions and CDF differences against state-updated
explicit sampling. The step scan and doubled-majorant control are reported
in the accompanying table; the curves compare the 50-step Gamma approximations.}
\end{figure}

\section{Electron-energy distributions and reference refinement}
This scan extends the distributional comparison beyond the central
$\chi_0=0.002$ state-updated test. Its references are refined frozen-packet
calculations, with their own step-refinement controls.
For all three initial quantum parameters, the final distributions are also
compared with independent explicit frozen-Poisson calculations using 4000 steps and $4\times10^4$ electrons.
Independent 2000-step controls contain $3\times10^4$ electrons. Their
Kolmogorov distances are 0.00425, 0.00438, and 0.00402 for increasing
$\chi_0$. The refined references agree at the resolution of these independent ensembles.

For the 50-step aggregate runs, the CDF distances are
\begin{center}
\begin{tabular}{crrr}\toprule
$\chi_0$ & Frozen Gamma & Generator correction & Midpoint/transport\\\midrule
0.002 & 0.08810 & 0.00573 & 0.00695\\
0.01  & 0.03620 & 0.00707 & 0.00649\\
0.02  & 0.02681 & 0.00380 & 0.00552\\\bottomrule
\end{tabular}
\end{center}
Across the nine generator-corrected field/step comparisons, the empirical
CDF distances range from 0.00252 to 0.00780. These values and the independent
reference-refinement distances specify the resolution of the comparisons;
they are not bounds on the error of the aggregate transition law.

\subsection{Quantile checks and uncertainty conventions}
Seven electron-energy quantiles are examined at probabilities
$0.001,0.01,0.05,0.5,0.95,0.99,0.999$. For each model and reference quantile,
a 97.5\% binomial order-statistic interval is formed. Subtracting the intervals
gives a conservative interval with at least 95\% pointwise coverage for their
difference, by a union bound. The coverage is pointwise in quantile and run. All 63 generator-corrected quantile intervals include zero.
The 50-step midpoint/transport case at $\chi_0=0.02$ has two marginal
exclusions, at the first and fifth percentiles, in this pointwise analysis. Both exclusions disappear in the finer-step runs.

At the outermost quantile the refined reference contains only about 40
particles in the tail, versus 60 in an aggregate ensemble. Accordingly the
sampled tail counts set the resolution of the electron-distribution test.
By $u=1$ the no-event atom is effectively absent on these sample sizes; its
weight is checked separately in the frozen/short-time analysis.

\begin{figure}[tbp]
\centering\includegraphics[width=\textwidth]{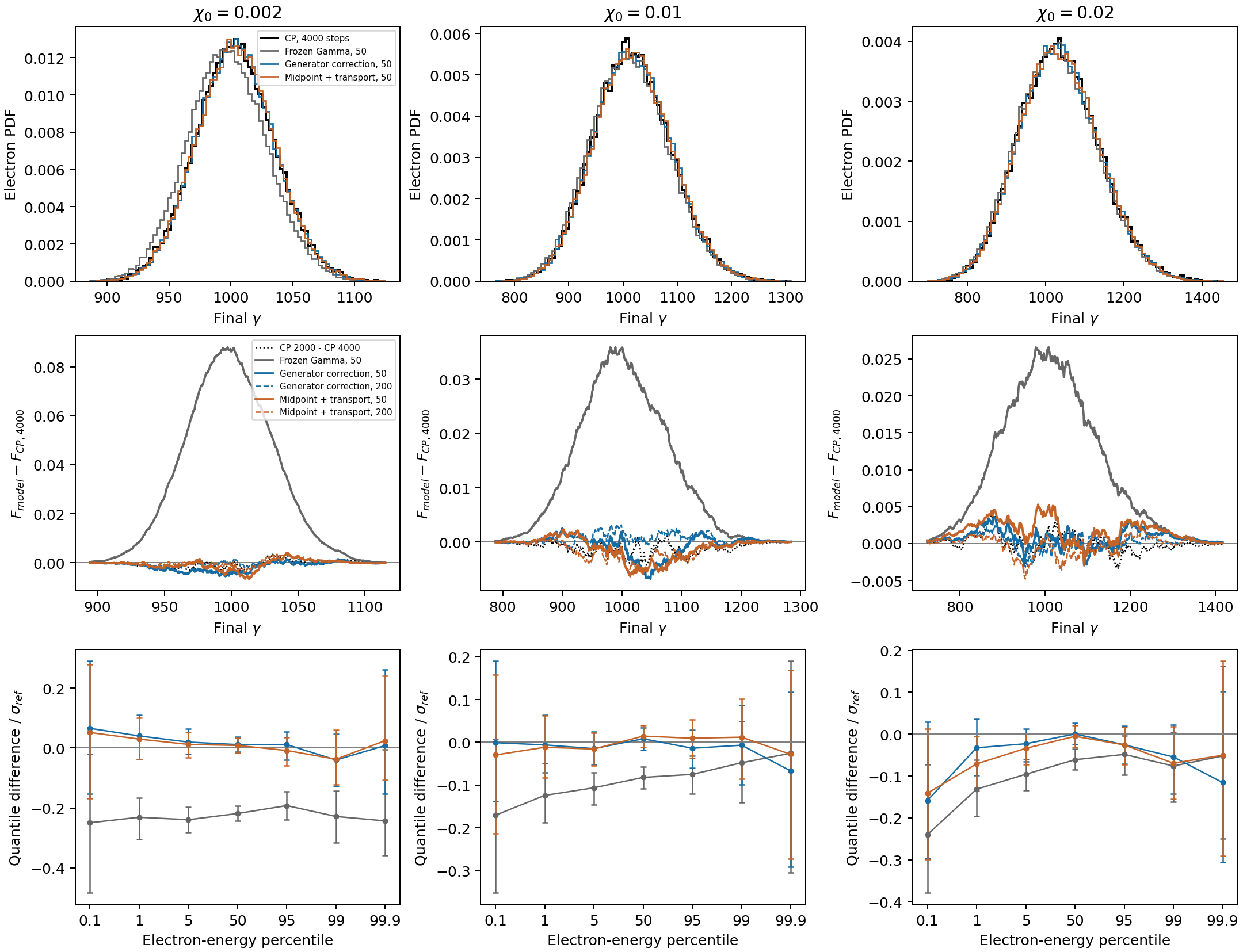}
\caption{Full energy-distribution comparison. Columns correspond to the three
initial quantum parameters. Top: final electron-energy histograms. Middle:
CDF differences from the 4000-step explicit reference, including an independent
2000-step reference check. Bottom: 50-step quantile differences normalized by
the refined-reference standard deviation, with conservative pointwise intervals.
The counts, grids and uncertainty conventions are stated in the text.}
\end{figure}

\section{Auxiliary midpoint/transport comparator}
A distinct soft-recoil comparator uses
\begin{equation}
 g_m=g-\tfrac12hA(g),\qquad
 M_m=hA(g_m),\qquad V_m=hB(g_m)[1-hA'(g_m)].
\end{equation}
The no-event probability still uses $W(g)$, not $W(g_m)$.
The variance factor captures cooling-induced contraction. A plain midpoint
substitution alone would omit this contribution. The connection to the
integral correction follows from a soft-recoil expansion:
\begin{align}
 \mathcal LA&=-AA'+\tfrac12BA''-\tfrac16R_3A'''+\cdots,\\
 \mathcal LB&=-AB'+\tfrac12BB''-\tfrac16R_3B'''+\cdots,\\
 C&=-BA'+\tfrac12R_3A''-\tfrac16R_4A'''+\cdots.
\end{align}
The midpoint/transport expressions retain dominant terms but omit others;
they define a distinct soft-recoil approximation to the integral correction.

\section{Magnetic coefficient evaluation and implementation checks}
The functions $A,B$ come from the checked moment table. The coefficients
$\mathcal LA,\mathcal LB,C$ are evaluated by direct quadrature of the continuum
CCF kernel with recoil-dependent $A,B$, using 384 Gauss--Legendre nodes in a
cubic transformed variable and truncating at transformed argument 80.
They are precomputed on 401 energies in $[400,2000]$ for each magnetic field
and interpolated with PCHIP. Doubling quadrature order at selected energies
changes the operator coefficients by at most approximately $1.3\times10^{-7}$
relatively. Selected off-grid interpolation checks have discrepancies up to
$6.93\times10^{-5}$. These checks quantify the selected quadrature and interpolation samples.
The continuum operator quadrature and interpolated local moments also inherit
the small table mismatch quantified above.

The auxiliary intermediate-state variant differentiates a PCHIP representation of $A$.
The run checks admissible moments, positive conditional variance, grid
coverage and the available-energy boundary at every update. A failed check
stops the run.
None occurred in these ensembles. The largest relative correction to the
frozen mean was about 0.020 and to the frozen variance about 0.078.
The coefficients are precomputed once per uniform magnetic field. The quadrature/PCHIP implementation here is the independent magnetic
verification implementation. Its one-time setup and per-step lookup should
not be identified with the portable C++ correction lookup and timings
described separately in Supplement I.

\section{Aggregation and orbit resolution}
The parameters are $b=B/B_{\rm cr}=10^{-6}$, $\gamma_0=1000$, $\mathbf p_0\parallel\mathbf e_x$, and $\mathbf B\parallel\mathbf e_z$. Define
\begin{equation}
 \omega_{B0}=b/(\tc\gamma_0),\qquad
 \rho_{B0}=\sqrt{\gamma_0^2-1}\,c\tc/b.
\end{equation}
The duration $\omega_{B0}T=10\pi$ is approximately 40.47 ps; $\rho_{B0}\simeq386\,\mu$m. There are $10^4$ electrons per case. The orbital step is $h_B=\omega_{B0}\Delta t_{\rm phys}=(b/\gamma_0)\Delta\tau$; this is distinct from the natural-time step $h=\Delta\tau$ in the cooling expansion. The initial quantum parameter is $\chi_0\simeq10^{-3}$. The 3950-step frozen explicit reference emits 330.505 photons per electron, with count standard deviation 17.947. Its mean photon energy is 0.20398 MeV. Of 3305050 photons, 197 carry more than 1\% of the pre-update electron energy and none more than 5\%.

\begin{table}[tbp]\centering\small
\caption{Magnetic scan. The phase error is for Gamma, relative to continuous rotation with the same recoil history. Multiplicity is measured from explicit frozen-Poisson sampling. No paths stop.}
\begin{tabular}{rrrrrr}\toprule
Steps & $h_B$ & $\langle\gamma\rangle_\Gamma$ & $\langle\gamma\rangle_{\rm P}$ & photons/nonempty & phase error (deg)\\\midrule
3950 & .00795 & --- &868.069&1.043&---\\
316&.09942&867.992&867.749&1.614&.232\\
158&.19883&867.813&867.741&2.391&.931\\
79&.39767&867.705&867.717&4.25&3.74\\
40&.78540&867.459&867.481&8.27&14.81\\
20&1.57080&867.212&866.992&16.54&63.38\\\bottomrule
\end{tabular}\end{table}

\subsection{Relativistic implicit-midpoint rotation}
For a Lorentz drift interval $h_L=\omega_{B0}\delta t$, the midpoint method preserves endpoint momentum magnitude. The Lorentz factor at the arithmetic midpoint momentum determines the exact
nonlinear rotation map.
Let $k=\gamma_0h_L/2$, $a=\gamma^2-k^2$, and $v=\tan^2(\vartheta/2)$. Then
\begin{equation}
 v^2+av-k^2=0,\qquad \vartheta=2\arctan\sqrt v,\qquad
 v=\begin{cases}2k^2/(\sqrt{a^2+4k^2}+a),&a\geq0,\\
 (\sqrt{a^2+4k^2}-a)/2,&a<0.\end{cases}
\end{equation}
This stable algebraic solution preserves the endpoint momentum norm. In $X=x/\rho_{B0}$, $Y=y/\rho_{B0}$, and $\mathbf U=\mathbf p/\sqrt{\gamma_0^2-1}$, the engine's electron-sign convention gives $\Delta X=U_y'-U_y$, $\Delta Y=-(U_x'-U_x)$.
The continuous rotation for the same recoil history in a full step is
\begin{equation}
 \Delta\Phi_{\rm history}=\frac{h_B}{2}\gamma_0
 (1/\gamma_{\rm before}+1/\gamma_{\rm after}).
\end{equation}
Subtracting its sum from the midpoint angles isolates the orbit-integration contribution.
This phase error concerns the trajectory. In this uniform field, rotation
phase does not change the emission rate or energy moments; finite-step
cooling shifts must therefore be distinguished from rotation error. The
orbit budget limits aggregation only through the common trajectory/radiation
step of this implementation. All 110000 magnetic trajectories complete, with relative energy-budget residuals below $7.5\times10^{-14}$.

A classical Landau--Lifshitz laboratory-time reference is
\begin{align}
 c_B&=\tfrac23\alpha b\gamma_0,\quad \zeta_0=\operatorname{atanh}(1/\gamma_0),\quad s_B=\omega_{B0}t,\\
 \gamma_{\rm LL}&=\coth(\zeta_0+c_Bs_B),\qquad u_{\rm LL}=\operatorname{csch}(\zeta_0+c_Bs_B),\\
 \Phi_{\rm LL}&=(\gamma_0/c_B)\log[\cosh(\zeta_0+c_Bs_B)/\cosh\zeta_0].
\end{align}
The final $\gamma_{\rm LL}=867.4267$ supplies a deterministic classical comparison;
the refined photon-sampled ensemble supplies the stochastic quantum reference.
Here $u_{\rm LL}$ denotes the momentum magnitude in units of $mc$.

\begin{figure}[tbp]
\centering\includegraphics[width=\textwidth]{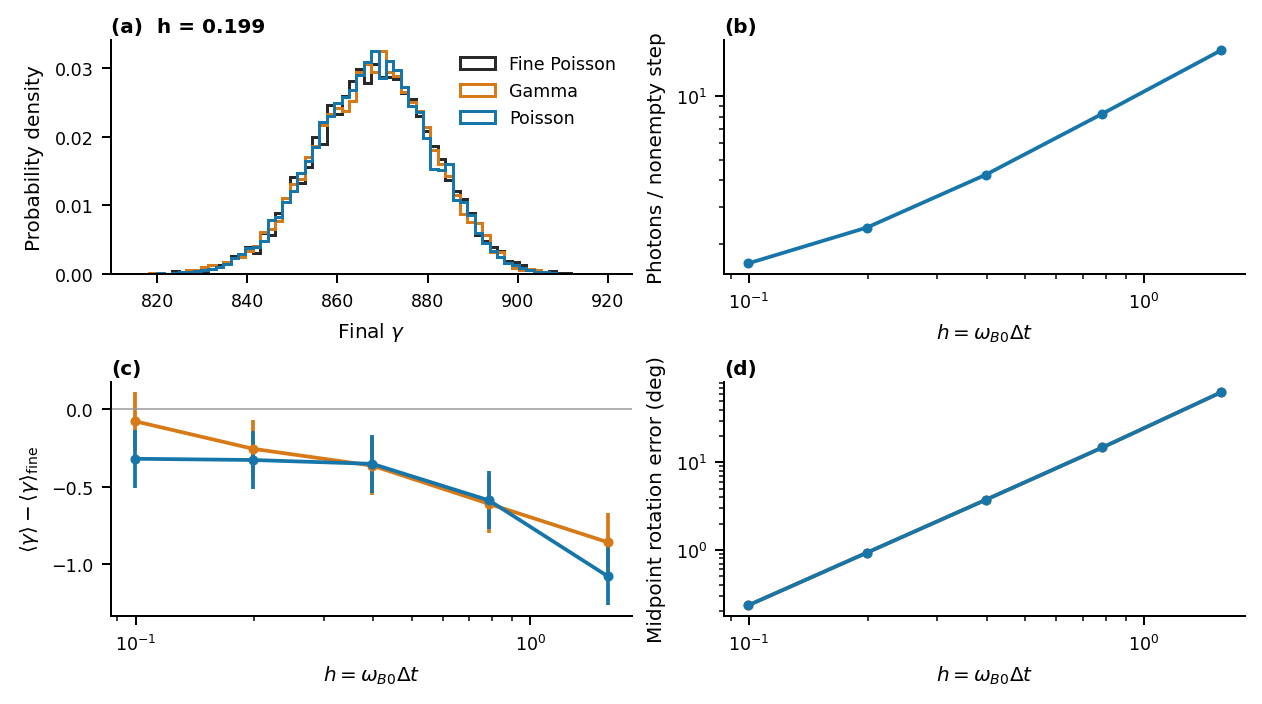}
\caption{Resolved-orbit magnetic benchmark. (a) Final energy distributions at
$\omega_{B0}\Delta t\simeq0.199$ versus a fine explicit reference.
(b) Photons per nonempty explicit step. (c) Mean-energy shifts from the fine
reference, with combined standard errors. (d) Pure midpoint rotation error
conditioned on the same recoil history. The last quantity isolates the orbit
integrator and is not a total trajectory-accuracy estimate.}
\end{figure}

\section{Trajectory-level diffusion comparison}
The magnetic support illustration tests the energy boundary separately
from the aggregation and analytical-cooling scans above. We set
$\mathbf E=0$, $\mathbf B=b\mathbf e_z$, $\gamma_0=1000$, and
$b=0.04/\sqrt{\gamma_0^2-1}\simeq4.000002\times10^{-5}$,
so that $\chi_0=0.04$ exactly, with initially transverse momentum.
Each model uses $N=4\times10^4$ electrons. Laboratory time is expressed as
$s_B=\omega_{B0}t$, where $\omega_{B0}=b/(\tc\gamma_0)$.
The interval $0<s_B\leq2\pi$ is resolved by 1024 external steps and
repeated with 2048 steps. The initial load is approximately 0.0625
photons per coarse step. This is a low-local-multiplicity comparison of
the recoil closure and its support, not the high-multiplicity aggregation test.

Let $A=Wm_1$ and $B=Wm_2$ denote the raw LCFA moment rates of
Supplement~I per unit $\tau=t/\tc$. The energy-space diffusion truncation is
\begin{equation}
 \dd\gamma=-A_B(\gamma)\dd s_B+\sqrt{B_B(\gamma)}\dd\mathcal W_{s_B},
 \qquad A_B=\frac{\gamma_0}{b}A,\quad B_B=\frac{\gamma_0}{b}B.
 \label{eq:langevin}
\end{equation}
Here $\mathcal W_{s_B}$ is a standard Wiener process. This is an It\^o
equation, with diffusion term $\tfrac12\partial_\gamma^2(B_Bf)$ in the
forward operator. In particular, $m_2$ is the raw second photon moment,
not its variance. Both coefficients use the same numerical LCFA table
as the positive samplers.

The scalar Milstein update \cite{Higham2001siam} is
\begin{equation}
 \gamma_{n+1}=\gamma_n-A_Bh_B+\sqrt{B_B}\,\Delta W
       +\tfrac14 B_B'(\Delta W^2-h_B),\qquad
 \Delta W\sim\mathcal N(0,h_B).
\end{equation}
Here $h_B=\Delta s_B$; coefficients and the derivative are evaluated at
$\gamma_n$. The derivative follows the piecewise log--linear coefficient
interpolants. No upper-energy clipping, reflection or redraw is used.
Corrected Gamma sampling uses the state-feedback moments of Supplement~I.
The state-updated explicit reference uses uniformization of the local
jump generator, recomputing the emitting state after each accepted photon
while keeping the external interval fixed.

Momentum is reconstructed on the mass shell as
$r=|\mathbf p|/p_0=\sqrt{(\gamma^2-1)/(\gamma_0^2-1)}$, with phase increment
$\Delta\phi=h_B\gamma_0(1/\gamma_n+1/\gamma_{n+1})/2$.
This couples collinear recoil to magnetic rotation without independent
angular diffusion. The energy-distribution comparison below does not
establish the accuracy of this phase approximation.

All histories complete on both grids, and neither positive sampler
exceeds $\gamma_0$ at any saved observation time. For the energy CDFs,
$D=\sup_\gamma|F_{\rm model}-F_{\rm updated}|$ gives
\begin{center}
\begin{tabular}{rrrrr}\toprule
&\multicolumn{2}{c}{Corrected Gamma}&\multicolumn{2}{c}{Langevin--Milstein}\\
$s_B$&1024&2048&1024&2048\\\midrule
$\pi/8$ &0.00945&0.01128&0.14548&0.14578\\
$\pi$   &0.00728&0.00353&0.04663&0.04845\\
$2\pi$  &0.00703&0.00495&0.03283&0.03318\\\bottomrule
\end{tabular}
\end{center}
At $s_B=\pi/8$, the fractions of Langevin electrons with
$\gamma>\gamma_0$ are 0.145475 and 0.145775 on the two grids; their
marginal binomial standard error is approximately 0.00176. These are
instantaneous fractions, not probabilities of ever crossing the boundary.
At $s_B=\pi$, the corresponding fractions are 0.001175 and 0.001125.
The early exterior population therefore persists under step refinement.

Same-model CDF distances between the two grids are at most 0.00813 over
the saved times. The grids reuse seeds but do not have nested Wiener
increments or a pathwise coupling; their differences include Monte Carlo
fluctuations. Likewise, the small Gamma--reference discrepancy is a
finite-sample measurement and need not be entirely sampling noise:
two-moment matching does not reproduce the single-photon spectrum exactly.
All CDFs and exterior fractions use initial-population normalization;
there is no discarded probability mass in these runs.

The upper panels of the main-text figure show six synchronous snapshots
at $s_B=\pi/8,\pi/4,3\pi/8,\pi/2,3\pi/4,\pi$.
Each continuous density is normalized by the initial $N$ and exact polar
cell areas, using nominal radial and angular bin widths 0.008; the radial
mesh includes $r=1$ exactly. The displayed color is their pointwise maximum,
not a summed density or time average. The no-emission atom is excluded
from the continuous density and marked separately. Gray histories are
the first three particle IDs. Three distinct completed red Langevin histories
are selected from all $4\times10^4$ electrons, excluding these ordinary IDs:
the largest maximum radius, the longest consecutive exterior interval, and
the largest time-integrated positive radial excess, in that order and excluding
previous selections. These grid-based criteria use the full saved interval
$0\leq s_B\leq2\pi$. Two deterministic replays identify the global IDs and
record the selected histories while preserving the original vector random
draws; the saved Milstein snapshots are reproduced exactly. These curves are
illustrative extrema, not a frequency sample. The lower panels
show energy CDFs at $s_B=\pi/8$ and $\pi$.

This comparison complements the frozen loss-law illustration in
Supplement~I: diffusion does not retain the support of the emission process,
whereas positive sampling preserves it. It does not exclude a useful
diffusion limit for observables insensitive to this distinction.

\section{Auxiliary diffusion conventions and frozen Gaussianization}
The photon energy is positive, while the electron energy jump is $\Delta\gamma=-\epsilon$. Hence the infinitesimal electron jump moments are $a_n=(-1)^nR_n$, and the formal energy-space Kramers--Moyal expansion is
\begin{equation}
 \partial_\tau F(\gamma,\tau)=\sum_{n=1}^\infty\frac1{n!}
 \partial_\gamma^n[R_n(\gamma)F(\gamma,\tau)].
\end{equation}
The drift is $A_\gamma=-R_1$. If the diffusion term is written as $\partial_\gamma^2(D_\gamma F)$, then $D_\gamma=R_2/2$; equivalently the It\^o noise amplitude is $\sqrt{R_2}$. With dimensionless momentum $\mathbf u=\mathbf p_{\rm phys}/(mc)$, strictly collinear photons have direction $\hat{\mathbf p}=\mathbf p_{\rm phys}/|\mathbf p_{\rm phys}|$ and positive recoil tensors $\mathcal R^{(n)}_{i_1\ldots i_n}\simeq R_n\hat p_{i_1}\cdots\hat p_{i_n}$. Replacing $\hat{\mathbf p}$ by $\mathbf v/c$ adds an ultrarelativistic approximation. These tensors describe the leading collinear jump; the finite mass-shell recoil map used in the simulations is specified in Supplement I.

The nonzero higher coefficients distinguish the emission generator from its
second-order diffusion closure. The diffusion limit and positivity of finite
Kramers--Moyal truncations are treated in
Refs.~\cite{Moyal1949,Pawula1967,Niel2018pre}. The coefficients in Eq.~\eqref{eq:soft_rates} are coefficients of the
constant-crossed-field kernel and therefore also apply to that field geometry.

For the frozen fractional cumulative loss $Z=S/\gamma=\sum_{j=1}^Ns_j$, $N\sim\operatorname{Poisson}(\Lambda)$, the exact compound-Poisson relation is
\begin{equation}
 \operatorname{cum}_n(Z)=\Lambda\mu_n,
\end{equation}
not $\Lambda$ times the cumulant of a single $s_j$. Thus, to leading order in $\chi$,
\begin{align}
 \operatorname{Skew}(Z)&=
 \frac{28\sqrt3/15}{(11/12)^{3/2}\sqrt\Lambda}[1+O(\chi)]
 =\frac{3.68392\ldots}{\sqrt\Lambda}[1+O(\chi)],\\
 \operatorname{ExKurt}(Z)&=
 \frac{1309/80}{(11/12)^2\Lambda}[1+O(\chi)]
 =\frac{19.47273\ldots}{\Lambda}[1+O(\chi)].
\end{align}
Small $\chi$ controls individual recoil size; large local $\Lambda$ controls Gaussianization of a frozen sum. Gaussianization is therefore a local property of the frozen emission statistics.

\section{Results and domain}
The synchrotron calculation fixes the single-photon moment normalization,
and the analytical energy moments provide an independent cooling reference.
Their second-order terms separate physical quantum corrections from the
finite-step cooling bias shared by frozen photon packets and frozen Gamma
sampling. Including recoil-state feedback suppresses that bias while retaining
aggregation; the state-updated explicit reference and three-$\chi_0$ distribution
scan test the resulting energy distributions.

The orbit scan separately shows that photon multiplicity can exceed unity
before trajectory error becomes restrictive. In the Langevin comparison,
the exterior population persists under step refinement while both positive
samplers preserve the initial-energy boundary, illustrating why agreement
of two local moments does not ensure the same physical support.

Together these checks support soft-photon aggregation as a stochastic
particle update. The Gamma construction retains the selected zero mass and
first two moments, and the feedback correction improves cooling without
individual photon draws. State variation, orbit resolution, distributional
shape and excessive-loss risk delimit the tested window.

\bibliographystyle{apsrev4-2}\bibliography{references}